\documentclass[sigconf]{acmart}
\AtBeginDocument{%
  }

\copyrightyear{2026}
\acmYear{2026}
\setcopyright{cc}
\setcctype{by}
\acmConference[UIST '26]{The 39th Annual ACM Symposium on User Interface Software and Technology}{November 02--05, 2026}{Detroit, MI, USA}
\acmBooktitle{The 39th Annual ACM Symposium on User Interface Software and Technology (UIST '26), November 02--05, 2026, Detroit, MI, USA}
\acmDOI{10.1145/3830398.3830693}
\acmISBN{979-8-4007-2856-3/2026/11}

\usepackage{soul}  % Used for custom comments
\usepackage{color} % Used for custom colors in comments
\usepackage{xspace}
\usepackage{listings}
\usepackage{enumitem}
\usepackage{multirow}
\usepackage{tabularx}   % 添加这个包
\newcommand{\eg}{{\it e.g.,\ }}

\usepackage{booktabs}
\definecolor{oxfordblue}{rgb}{0.0, 0.13, 0.28}
\definecolor{harvardcrimson}{rgb}{0.79, 0.0, 0.09}
\definecolor{dartmouthgreen}{rgb}{0.05, 0.5, 0.06}
\definecolor{princetonorange}{rgb}{1.0, 0.56, 0.0}
\definecolor{yaleblue}{rgb}{0.06, 0.3, 0.57}
\definecolor{usccardinal}{rgb}{0.6, 0.0, 0.0}
\definecolor{uclablue}{rgb}{0.33, 0.41, 0.58}
\definecolor{msugreen}{rgb}{0.09, 0.27, 0.23}
\definecolor{cornellred}{rgb}{0.7, 0.11, 0.11}
\definecolor{pomegranate}{RGB}{192, 57, 43}
\definecolor{anti-pomegranate}{RGB}{43,178,192}
\definecolor{alizarin}{RGB}{231, 76, 60}
\definecolor{anti-belize}{RGB}{185, 41, 56}
\definecolor{belize}{RGB}{41, 128, 185}
\definecolor{peter}{RGB}{52, 152, 219}
\definecolor{green}{RGB}{22, 160, 133}
\definecolor{anti-green}{RGB}{160,22,118}
\definecolor{turquoise}{RGB}{26, 188, 156}
\definecolor{pumpkin}{RGB}{211, 84, 0}
\definecolor{anti-pumpkin}{RGB}{0,22,211}
\definecolor{carrot}{RGB}{230, 126, 34}
\definecolor{wisteria}{RGB}{142, 68, 173}
\definecolor{anti-wisteria}{RGB}{99,173,68}
\definecolor{amethyst}{RGB}{155, 89, 182}
\definecolor{nephritis}{RGB}{39, 174, 96}
\definecolor{anti-nephritis}{RGB}{174,39,117}

\newcommand{\haoxiang}[1]{{{#1}}}
\newcommand{\zhh}[1]{{#1}}
\newcommand{\zhhred}[1]{{#1}}

\newcommand{\pzh}[1]{{#1}}

\newcommand{\zzh}[1]{{#1}}
\newcommand{\fhx}[1]{{#1}}
\newcommand{\fan}[1]{{#1}}

\newcommand{\name}{{\textit{TeachUp}}}
    
\begin{document}

% \title{TeachUp: Making Learning Instructional Strategies from Videos More Reflective and Practice-Oriented for Teachers}
% \title{TeachUp: Supporting Teachers to Learn Instructional Strategies from Classroom Teaching Videos with Reflective Hints and Feedback}

% \title{\haoxiang{TeachUp: Supporting Novice Teachers to Learn Instructional Strategies from Classroom Teaching Videos with Reflective Hints and Feedback}}
\title{
TeachUp: Facilitating Early-Stage Teachers to Learn Instructional Strategies from Classroom Videos with Reflective Support
}

\author{Haoxiang Fan}
\orcid{0009-0000-5729-8491}
\affiliation{%
  \department{School of Artificial Intelligence}
  \institution{Sun Yat-sen University}
  \city{Zhuhai}
  \country{China}
}
\email{fanhx6@mail2.sysu.edu.cn}

\author{Ding Lei}
\orcid{0009-0004-4812-2135}
\affiliation{%
  \department{Institute of Artificial Intelligence and Brain Sciences}
  \institution{University of Macau}
  \city{Macau}
  \country{China}
}
\email{mc56574@um.edu.mo}

\author{Zaihong Zheng}
\authornote{These authors contributed equally to this work.}
\orcid{0009-0001-4078-1900}
\affiliation{%
  \department{School of Artificial Intelligence}
  \institution{Sun Yat-sen University}
  \city{Zhuhai}
  \country{China}
}
\email{cheng256@mail2.sysu.edu.cn}

\author{Jiale Li}
\authornotemark[1]
\orcid{0009-0002-4666-7700}
\affiliation{%
  \department{School of Artificial Intelligence}
  \institution{Sun Yat-sen University}
  \city{Zhuhai}
  \country{China}
}
\email{lijle27@mail2.sysu.edu.cn}

\author{Jionghao Lin}
\orcid{0000-0003-3320-3907}
\affiliation{%
  \institution{The University of Hong Kong}
  \city{Hong Kong}
  \country{China}
}
\email{jionghao@hku.hk}

\author{Jian Yin}
\orcid{0000-0002-1214-5384}
\affiliation{%
  \department{School of Artificial Intelligence}
  \institution{Sun Yat-sen University}
  \city{Zhuhai}
  \country{China}
}
\email{issjyin@mail.sysu.edu.cn}

\author{Zhenhui Peng}
\authornote{Corresponding author.}
\orcid{0000-0002-5700-3136}
\affiliation{%
  \department{School of Artificial Intelligence}
  \institution{Sun Yat-sen University}
  \city{Zhuhai}
  \country{China}
}
\email{pengzhh29@mail.sysu.edu.cn}

\renewcommand{\shortauthors}{Fan et al.}
\renewcommand{\shorttitle}{TeachUp}

\begin{abstract}
% Novice teachers can learn instructional strategies, e.g., how to activate students' prior knowledge, by watching recorded videos of others' offline classes. However, this video-based learning process can be challenging due to the pedagogical complexity of classroom teaching videos , which often demonstrate multiple strategies without clear guidance, and lack of reflection support, as revealed in our formative study (N=9). In this paper, we first develop an LLM-powered computational pipeline to detect nine teaching strategies in the videos. Then, we present TeachUp, an interactive system that supports teachers to learn teaching strategies by providing adaptive questions, hints, and feedback when watching videos and practicing. Our within-subjects study (N=16), comparing the system with traditional video-playing and self-practicing baseline, along with four teacher interviews, demonstrate that TeachUp is effective in improving the outcomes of learning instructional strategies in VBL and strengthening teachers' confidence in classroom instruction.

\zhh{
% Offline open-class teaching videos provide good examples for novices to learn instructional strategies, e.g., how to organize cooperative learning. 
Recorded videos of offline open classes provide good examples for early-stage teachers to learn instructional strategies, e.g., how to organize cooperative learning. 
However, learning by watching these videos is challenging, as these strategies are implicitly performed, and it lacks in-situ reflective support. 
% as novices have to recognize the performing instructional strategies and lack in-situ reflective support. 
In this paper, via a formative study (N=9), we design \name{} to support the learning of instructional strategies from classroom teaching videos.
\name{} adopts an LLM-powered pipeline to detect nine instructional strategies in videos (precision = 63.4\%), provides reflective questions and hints while watching, and generates customized practices with reflective feedback.  
A within-subjects study (N=16) shows that compared to a traditional video-playing and self-practicing baseline, early-stage teachers with \name{} are more engaged in learning and perform better in applying learned strategies to new tasks. 
Interviews with four in-service teachers further generalize our findings and \name{}'s use cases. 
We discuss practical implications for fostering video-based learning of instructional strategies. 
}

% Novice teachers can learn instructional strategies, e.g., how to activate students' prior knowledge, by watching recorded videos of others' offline classes. 
% However, this learning process can be challenging due to the pedagogical complexity of classroom teaching, which often demonstrates multiple strategies associated with subjects, and a lack of reflection support, as revealed in our formative study (N=9). 
% % In this paper, we first develop an LLM-powered computational pipeline to detect nine instructional strategies in the videos. 
% % Then, we present TeachUp, an interactive system that supports teachers to learn the implementation of instructional strategies by providing adaptive questions, hints, and feedback when watching videos and practicing. 
% \haoxiang{In this paper, we present TeachUp, an interactive system that automatically detects nine instructional strategies in classroom videos and supports teachers to learn their implementation by providing adaptive questions, hints, and feedback when watching videos and practicing.}
% Our within-subjects study (N=16) using a traditional video-playing and self-practicing baseline,
% along with four teacher interviews, demonstrates TeachUp's effectiveness in promoting novice teachers' reflection and enhancing their ability to apply learned instructional strategies.
% We discuss practical implications for fostering teachers' professional development via video-based learning. 

\end{abstract}




\maketitle

\section{Introduction}
\zhh{
Instructional strategies, such as cooperative learning, setting objectives, and providing feedback ~\cite{marzano2003works}, are useful for enhancing student achievement in  offline classrooms. 
This paper targets \textbf{early-stage teachers}, including pre-service teachers (\eg students in normal universities) and in-service teachers with no more than one year of teaching experience, 
% those under training (\eg students in normal universities) and new in-service teachers (\eg with less one year of experience), 
and their learning of instructional strategies with classroom teaching videos. 
Specifically, the excellent open classes like the award-winning ones are usually recorded and available on video-sharing platforms.
% ~\footnote{\zhhred{\url{https://www.bilibili.com}}
These videos serve as valuable learning materials, in which well-trained teachers demonstrate good practices of instructional strategies in a real classroom. 
Besides, such video-based learning (VBL) has been increasingly adopted in professional development programs for both pre-service and in-service teachers~\cite{gaudin2015video, sablic2021video, leblanc2018analysis}.
}
\fan{Teacher-oriented VBL includes observing other teachers' classroom videos and reviewing one's own teaching videos~\cite{gaudin2015video}, and this work focuses on the former.}
% \pzh{
% \fhx{
% Teachers need to employ effective strategies, such as facilitating cooperative learning and engaging students in generating and testing hypotheses~\cite{marzano2003works}, to enhance student achievement in the offline classroom.
% }
% % Teachers in offline open classes usually showcase these teaching strategies in normally 40-minutes lecture sessions with real human students [refs].
% The excellent classes, \eg the award-winning ones in teaching competitions, are often video-recorded and shared in online platforms like Bilibili~\footnote{\url{https://www.bilibili.com}}, making them accessible for users to learn how to implement instructional strategies by watching videos.
% % Such a video-based learning (VBL) approach has been increasingly adopted in professional development programs for both pre-service and in-service teachers, spanning primary and middle schools across various countries~\cite{gaudin2015video, sablic2021video}.
% % By watching videos of well-prepared classes, teachers could observe teaching practices and learn skills from others~\cite{gaudin2015video, sablic2021video, leblanc2018analysis}.
% \haoxiang{Such video-based learning (VBL) has been increasingly adopted in professional development programs, enabling both pre-service and in-service teachers to observe teaching practices and learn skills from others~\cite{gaudin2015video, sablic2021video, leblanc2018analysis}.}
% }

\zhh{
% Despite the benefits, learning instructional strategies with such classroom videos is challenging, especially for novice teachers who are under training (\eg students in normal universities) or have little teaching experience (\eg less than three years).
% Despite the benefits, learning instructional strategies from classroom videos can be challenging for early-stage teachers.
% teachers---those still under training (\eg students in normal universities) or with little teaching experience (\eg less than three years)---who are the primary target users of our work.
Despite the benefits, learning instructional strategies from classroom videos is challenging for early-stage teachers.
\fan{Unlike directly replicable actions, an instructional strategy is defined by when and why it is enacted: because pedagogical content knowledge integrates pedagogy with disciplinary
content~\cite{shulman1986those, berry2016pedagogical}, the same strategy can be enacted differently across subjects and student states.
Yet classroom videos mainly expose teachers' observable actions and leave the underlying pedagogical reasoning implicit~\cite{spiro2014reflections}, which brings two challenges.}
First, these teachers have to recognize and analyze the instructional strategies performed by the teachers in the videos, while they usually lack teaching experience using these strategies. 
Recorded videos of open classes are often pedagogically complex, because they feature teachers effectively demonstrating multiple strategies, yet lack the descriptions and tags needed to navigate the video content based on instructional strategies~\cite{amador2020video}.
% \fan{First, early-stage teachers need to locate instructional strategies in pedagogically complex classroom videos. 
% Such recordings often demonstrate multiple strategies but lack strategy-level descriptions or tags that would help teachers navigate the content~\cite{amador2020video}.}
% As a result, the mentally demanding and time-consuming process of identifying and analyzing these strategies often leads early-stage teachers to avoid longer, more complex videos, causing potentially high-value pedagogical examples to be ignored~\cite{sablic2021video, amador2020video}.
\fan{As a result, this mentally demanding and time-consuming process often leads early-stage teachers to avoid longer, more complex videos, causing potentially high-value pedagogical examples to be ignored~\cite{sablic2021video, amador2020video}.}
% Previous HCI work has computationally modeled student engagement~\cite{sabuncuoglu2023developing, guo2014video}, student emotions~\cite{zeng2020emotioncues}, and tutorial videos~\cite{kim2014crowdsourcing}.
% However, little work has attempted to automatically detect the instructional strategies that are used to teach the content to the students.
% \haoxiang{However, little work has attempted to automatically detect instructional strategies in the videos, nor to provide interactive support for teachers to learn from them.}
% Although prior HCI work has computationally modeled student engagement~\cite{sabuncuoglu2023developing, guo2014video} and emotions~\cite{zeng2020emotioncues} in classroom videos, 
% little work has attempted to address the following research question: \textbf{RQ1) How to computationally identify the instructional strategies of teachers from classroom teaching videos?}
% little work has attempted to automatically detect instructional strategies in classroom videos, nor to provide interactive support for novice teachers to learn from them.
}
\zhh{
% Secondly, early-stage teachers lack in-situ reflective support and practices when watching the classroom videos, which are essential in the learning task. 
\fan{Second, early-stage teachers lack in-situ reflective support while watching classroom videos and opportunities to practice the observed strategies afterward, both of which are essential to learning instructional strategies.}
% In fact, many teachers are prone to copying the teaching activities in the videos and lack engagement in reflection during VBL, avoiding activities like interpretation and evaluation~\cite{bates2016if}.
% This makes them miss crucial opportunities for professional growth.
\fan{In fact, many teachers are prone to copying the teaching activities in the videos rather than interpreting and evaluating them during VBL~\cite{bates2016if}, missing crucial opportunities for professional growth.}
% \fan{More fundamentally, learning instructional strategies requires teachers to interpret their situated use and consider how to transfer them to their own classrooms~\cite{spiro2014reflections, hassler2020oer4schools, stigler2009teaching}. Because pedagogical content knowledge integrates pedagogy with disciplinary content, the same strategy may be enacted differently across subjects~\cite{shulman1986those, berry2016pedagogical}.
% Yet classroom videos primarily show observable actions, leaving early-stage teachers to infer the underlying pedagogical reasoning rather than merely imitate what they see.}
Moreover, improving teaching skills depends heavily on practices in scenarios close to real classroom settings~\cite{hassler2020oer4schools, stigler2009teaching}.
% Yet, there is a lack of support for practicing the learned teaching strategies after watching the videos.
% Existing intelligent systems have adopted reflective hints and adaptive performance feedback to train teachers in simulated scenarios like one-on-one tutoring~\cite{wang2021practice, pan2025tutorup} and parent-teacher conversations~\cite{thompson2019teacher}.
% Nevertheless, these previous systems usually focus on text-based scenario or rely on pre-scripted, limited video-based scenarios.
% Yet, existing intelligent systems that train teachers in simulated scenarios~\cite{wang2021practice, pan2025tutorup, thompson2019teacher} usually focus on text-based or pre-scripted settings, lacking support for practicing strategies learned from diverse classroom videos.
}

\fan{Yet existing systems fall short on both sides.}
% VBL systems for other skills, \eg surgical procedures~\cite{nazari2020one}, public speaking~\cite{wang2020voicecoach}, and running~\cite{liu2022posecoach}, assume predefined target skills and observable performance dimensions, and thus do not support interpreting why a strategy is used in a situated context. Meanwhile, systems that train teachers in simulated scenarios~\cite{wang2021practice, pan2025tutorup, thompson2019teacher} focus on text-based or pre-scripted settings, without connecting practice to the strategies observed in diverse classroom videos.
\fan{
% For the first, video annotation tools scaffold teachers to mark and structure their reflection on classroom recordings~\cite{rich2009video}, and computational approaches summarize classroom videos by student emotions~\cite{zeng2020emotioncues} or engagement~\cite{sabuncuoglu2023developing}; neither surfaces which instructional strategy a teacher is enacting, or why it fits the classroom conditions at that moment.
For the first, video annotation tools help teachers structure reflection on classroom recordings~\cite{rich2009video}, and computational approaches summarize classroom videos by student emotions~\cite{zeng2020emotioncues} or engagement~\cite{sabuncuoglu2023developing}; neither surfaces which strategy a teacher is enacting, nor why it fits that classroom moment.
% For the second, VBL systems for other skills, \eg surgical procedures~\cite{nazari2020one}, public speaking~\cite{wang2020voicecoach}, and running~\cite{liu2022posecoach}, assume predefined target skills and observable performance dimensions, and thus do not support rehearsing a strategy whose enactment depends on the subject, the students, and the timing. Meanwhile, systems that train teachers in simulated scenarios~\cite{wang2021practice, pan2025tutorup, thompson2019teacher} focus on text-based or pre-scripted settings, without connecting practice to the strategies observed in diverse classroom videos.
For the second, VBL systems for other skills, \eg surgical procedures~\cite{nazari2020one}, public speaking~\cite{wang2020voicecoach}, and running~\cite{liu2022posecoach}, assume predefined skills with observable performance dimensions, and thus do not support rehearsing a strategy whose enactment depends on the subject, the students, and the timing. Scenario-based teacher-training systems~\cite{wang2021practice, pan2025tutorup, thompson2019teacher}, meanwhile, are text-based or pre-scripted, and do not connect practice to the strategies just observed in a video.
}
\fan{Given these two challenges, we ask the following research question: \textbf{How to design a reflective interaction loop to make early-stage teachers' learning of instructional strategies from classroom videos more effective and engaging?}}
% A remaining question is: \textbf{RQ2) How to support effective and engaging learning of instructional strategies from classroom videos for early-stage teachers?}
% It remains largely under-investigated how to support teachers to effectively learn instructional strategies of their interest from diverse offline classroom videos.

\zhh{
In this paper, we propose \name{}, an interactive system that supports early-stage teachers to learn instructional strategies based on offline classroom videos. 
We followed a user-centered design approach to develop and evaluate \name{}.
Initially, we interviewed four early-stage teachers to understand their practices and challenges of learning to teach from classroom videos. 
We also interviewed five experienced teachers to gather feedback and requirements for \name{}'s prototype from experts' views.
% a formative study involving six novice teachers and three experienced teachers who were responsible for programs of teachers' professional development.
\zhh{
The findings highlight the needs to deepen reflection on instructional strategies while watching the videos and to facilitate practicing these strategies in authentic scenarios after watching. 
% The findings on the challenges of traditional VBL revealed a need not only to deepen reflection on teaching strategies but also to facilitate the practice of these strategies in authentic scenarios.
}
% To address these needs, we first developed a computational pipeline that automatically extracts video clips where teachers demonstrate one of nine effective teaching strategies, as classified by Marzano~\cite{marzano2001classroom}, achieving an overall precision of 63.4\%.
% Built on the detected instructional strategies in the videos, and supported by the literature on \peng{VBL systems for teachers~\cite{gaudin2015video, sablic2021video, leblanc2018analysis, baecher2018facilitating, major2018using, yousef2014state, weng2023competency} and theories of teacher education~\cite{shulman1986those, berry2016pedagogical, hokanson2004levels},}
% we designed and implemented \name{} that supports users in three stages of VBL. 
% \haoxiang{To address these needs, supported by the literature on \peng{VBL systems for teachers~\cite{gaudin2015video, sablic2021video, leblanc2018analysis, baecher2018facilitating, major2018using, yousef2014state, weng2023competency} and theories of teacher education~\cite{shulman1986those, berry2016pedagogical, hokanson2004levels},}
% we designed and implemented \name{} that automatically detects nine effective teaching strategies as classified by Marzano~\cite{marzano2001classroom} from classroom videos, and supports users in three stages of VBL.}
\zhh{
To address these needs, 
% supported by the literature on VBL systems for teachers~\cite{gaudin2015video, sablic2021video, leblanc2018analysis, baecher2018facilitating, major2018using, yousef2014state, weng2023competency} 
we designed and implemented \name{} based on the interview findings and theories of teacher education~\cite{shulman1986those, berry2016pedagogical, hokanson2004levels}. %, 
% we designed and implemented \name{}, which supports novice teachers in three stages of VBL.
% \name{} addresses RQ1 via an LLM-powered computational pipeline that detects nine instructional strategies suggested by \citet{marzano2013marzano} based on the video transcript, with an average precision of 0.634. 
% Then, in response to RQ2, \name{} supports the learning of each detected strategy in three stages.  
\fan{
% To address our research question,
\name{} structures reflection and practice into a three-stage interaction loop. To help users locate relevant examples, an LLM-powered pipeline suggests clips containing nine instructional-strategy categories defined by \citet{marzano2001classroom} (average precision: 0.634).}
}
\zhh{
While watching the videos, users receive generated reflective questions on the teachers' behaviors and alternative ways to enact a strategy.
Subsequently, \name{} provides a generated scenario, tasks as hints for users to practice the learned strategy by speaking to the camera, similar to microteaching training~\cite{allen1972microteaching}.
Finally, users can get a personalized report generated by a multi-modal large language model, which helps reflection by contrasting the user's performance with that in the watched video.
}
}

\zhh{
To evaluate \name{}'s effectiveness and user experience, we conducted a within-subjects study with 16 early-stage teachers, using a baseline system that supports conventional video watching and self-practicing in front of a camera. 
% We evaluated \name{} via two studies focusing on two research questions, \ie the usability and usefulness of \name{} (RQ1), and its effectiveness (RQ2). 
% One study is a within-subjects, controlled laboratory study with 16 novice teachers using a baseline system that supports conventional video watching and self-practicing in front of a camera. 
% The other is an interview study with four in-service teachers who freely used \name{} to learn any instruction strategy in videos of their specialized subjects. 
% in a within-subjects, controlled laboratory study ($N = 16$ novice teachers), focusing on two research questions: the usability and usefulness of \name{} (RQ1), and its effectiveness (RQ2).
% The quantitative results indicated that compared to the baseline condition, \name{} significantly improves participants' performance in applying the learned strategy after the learning session. 
% \name{} also significantly increases users' willingness to engage in VBL and teaching confidence, and participants perceived its supported practices significantly more useful. 
\zhh{Results showed that \name{} significantly improves participants' performance in applying the learned strategy from the video to a new teaching task. 
Users with \name{} reported being significantly more engaged in the learning session and increased teaching confidence. %, and is perceived as providing more useful practice support.
}
% users' willingness to learn in a video-based environment, that the generated practice hints are useful, and that it significantly enhances novice teachers' ability to apply strategies.
% We further gathered in-depth insights via interviews with two novice teachers and two experienced teachers who freely used \name{} to learn any instruction strategy in videos of their specialized subjects.
Participants especially praised the adaptive support of reflections during the watching, practicing, and evaluation stages. 
We further gathered in-depth insights via interviews with two early-stage teachers and two experienced teachers who freely used \name{} to learn any instruction strategy in videos of their specialized subjects. 
They indicated that the LLM-generated content in \name{} was overall helpful but sometimes distracting. 
Together, these findings suggest that \name{} could support teachers to learn from others' teaching videos effectively for professional development at scale. % VBL, especially the novice teachers.
% The qualitative findings suggested that the LLM-generated content in \name{} was overall helpful but sometimes distracting. 
% Participants especially praised the adaptive support of reflections during all the watching, practicing, and evaluation stages, suggesting a promising future for deploying \name{} to benefit a large scale of teachers in VBL, especially the novice teachers. 
% The findings demonstrate the usefulness of our instructional strategy detection pipeline and suggest that novice teachers can benefit from the ``watching–practicing–evaluating'' process.
}

\zhh{
In summary, this paper has three contributions.
\fan{First, we contribute the design and implementation of a reflective interaction loop that integrates watching, microteaching, and comparison-based feedback to help early-stage teachers locate, transfer, rehearse, compare, and revise instructional strategies learned from classroom videos.
% Second, via a within-subjects study, we demonstrate \name{}'s usefulness in improving learning engagement and performance in applying the learned strategy to new teaching tasks
Second, via a within-subjects study, we demonstrate \name{}'s usefulness in improving learning engagement, strategy noticing, and performance in applying the learned strategy to new teaching tasks. 
Third, as an enabling component, we contribute the first benchmark for instructional strategy detection in classroom videos and a preliminary computational pipeline, intended as starting points for future research.}\footnote{We provide examples, prompts, and dataset in supplementary materials.
}
}

\section{Related Work}
\label{sec:related_work}
\subsection{Video-Based Learning for Teachers' Professional Development}
% Video-Based Learning (VBL) emphasizes learning through watching videos~\cite{sablic2021video}, which not only supports student learning but also provides valuable opportunities for teachers to reflect in their professional development~\cite{gaudin2015video}. 
% These videos can be categorized into two types: videos of other teachers (\ie unknown teachers' videos, colleagues' teaching videos) and self teaching videos~\cite{zhang2010using}. 
% In the past 20 years, VBL has been increasingly integrated into teacher education~\cite{gaudin2015video, sablic2021video, tucholka2025analysing}. Various empirical studies have examined its effectiveness in professional development (PD) programs~\cite{gaudin2015video} and online self-learning~\cite{bates2016if} to improve the competencies of novice teachers, and summarized a range of valuable insights.
\zhh{
% Video-Based Learning (VBL) emphasizes learning through watching videos~\cite{sablic2021video} and provides valuable opportunities for teachers to reflect in their professional development~\cite{gaudin2015video}. 
Over the past 20 years, video-based learning (VBL) --- including watching videos of other teachers as well as self-teaching videos~\cite{zhang2010using} --- has been increasingly integrated into teacher education~\cite{gaudin2015video, sablic2021video, tucholka2025analysing}, with empirical studies examining its effectiveness in professional development (PD) programs~\cite{gaudin2015video} and online self-learning~\cite{bates2016if}.}
For example, \textit{STeLLA} provides the Lesson Analysis Protocol (LAP), a tool that helps teachers analyze classroom videos through a \textit{claim–evidence–reasoning–alternative} process focused on science content storyline, to facilitate reflection and critical thinking during PD programs~\cite{taylor2017effect}, which we integrate into \name{}.
\zhh{
When watching these videos, teachers tend to focus on general teaching moves~\cite{baecher2018facilitating} and, more specifically, on learning how to implement instructional strategies~\cite{siry2014facilitating, gaudin2015video}. 
However, the effectiveness of VBL remains debated~\cite{so2012quality, zhang2011understanding}. Two key reasons have been identified: the sheer volume and pedagogical complexity of available videos makes it hard for early-stage teachers to locate useful clips~\cite{amador2020video, sablic2021video}, and the absence of reflective thinking --- \eg evaluating video quality~\cite{so2012quality, zhang2011understanding} and considering how to transfer strategies to one's own classroom~\cite{hassler2020oer4schools, stigler2009teaching} --- limits competence development. 
% In HCI, few studies have explored interactive systems that address these gaps, which \name{} aims to tackle.
}

\fan{Beyond teacher education, VBL has supported training in surgical procedures~\cite{nazari2020one}, voice modulation in public speaking through exemplar matching~\cite{wang2020voicecoach}, and running technique through novice--expert pose comparison~\cite{liu2022posecoach}.
These systems rely on predefined target skills and observable performance dimensions, and they do not address a central need in teacher VBL: locating implicit strategies in complex classroom videos and interpreting their context-dependent rationale to guide transfer, rehearsal, and revision.}
\fan{Our work is motivated by the benefits and gaps of VBL for developing teaching competence and developing \name{} with strategy detection and reflective support to facilitate VBL of instructional strategies.}

\subsection{Analyses of Classroom Videos}
% To extract information from videos to support skill acquisition, researchers have explored various computational methods. 
% For instance, prior work has used crowdsourcing to extract the step-by-step macro-structure from how-to videos, enabling the creation of interactive players with easier navigation~\cite{kim2014crowdsourcing}.
% With the development of multimodal analysis techniques, recent studies utilize models to extract fine-grained features from video content.
% For example, \textit{VoiceCoach} ~\cite{wang2020voicecoach}
% analyzed thousands of high-quality talks to build an evidence base of voice modulation examples and, by tightly coupling multimodal analysis with formative feedback, provides immediate, quantitative visual guidance that scaffolds skill acquisition and practice. Inspired by this work, we also use such multimodal techniques to evaluate teachers' performance during their practice.
\zhh{Prior work has extracted actionable structure from videos for skill acquisition, from crowdsourced  segmentation in how-to videos~\cite{kim2014crowdsourcing} to multimodal analysis for formative feedback~\cite{wang2020voicecoach}.}
% In the field of education, video analysis techniques have been applied to support teachers' reflection, instruction, and classroom management. 
% They have been used to help teachers better understand classroom affective dynamics, provide timely interventions in both online and offline settings~\cite{zeng2020emotioncues, ma2022glancee}, 
% and facilitate student teamwork and evaluating students' outcomes~\cite{echeverria2024teamslides, zhang2025cpvis}. Notably, \textit{ClassInsight}~\cite{ngoon2024classinsight} supports teachers' personalized reflection by visualizing classroom discussion data.
% Instead of relying on videos, which require considerable effort, it uses transcripts with further processing, suggesting that classroom transcripts alone have the potential to support applications for effective teacher professional development.
\zhh{In education, such analysis has supported reflection and orchestration by surfacing classroom affective dynamics, online learning status, teamwork processes, and discussion patterns~\cite{zeng2020emotioncues, ma2022glancee, echeverria2024teamslides, zhang2025cpvis, ngoon2024classinsight}. 
\zhhred{For example, \citet{zeng2020emotioncues} help teachers better understand classroom affective dynamics, provide timely interventions in both online and offline settings via in-class video analysis. \textit{Glancee}~\cite{ma2022glancee} provides an adaptable system for instructors to grasp 
student learning status in synchronous online classes.}
\zhhred{Additionally, 
~\citet{ngoon2024classinsight} support teachers' personalized reflection by visualizing classroom discussion data.
Instead of relying on videos, which require considerable effort, it uses transcripts with further processing, suggesting that classroom transcripts alone have the potential to support applications for effective teacher professional development.
}
\zhh{
Nevertheless, there exists a gap in computationally identifying instructional strategies in the classroom videos and analyzing learners' performances of practicing these strategies. 
We develop an LLM-powered computational pipeline to detect instructional strategies based on video scripts and prompt multi-modal large models to evaluate learners' performances in practicing videos.
}
% \textit{ClassInsight} further shows that transcript-based processing alone can support personalized teacher reflection~\cite{ngoon2024classinsight}, which motivates our transcript-first design for scalable processing.
}

% Most existing tools focus on reflecting upon a teacher's own classroom recordings and lack an analytical approach applicable to a broader range of videos (\eg other teachers' classroom videos) for learning. An effective analytical approach could make teacher training more scalable, as it can leverage vast video resources. Our pipeline focuses on how to identify and analyze instructional strategies within classroom videos.
% \haoxiang{Most existing tools focus on a teacher's own classroom recordings and lack analytical approaches applicable to the broader range of videos available for learning. Our pipeline addresses this gap by identifying and analyzing instructional strategies within classroom videos, enabling more scalable teacher training.}

\subsection{Reflective Support Tools for Learners}
% Reflective learning is the process of internally examining and exploring an issue of concern, triggered by an experience, which helps individuals make sense of what happened, gain a deeper understanding of themselves, and develop new perspectives~\cite{boyd1983reflective}.
% In the field of HCI, various research has focused on designing interactive systems that promote reflective learning~\cite{bentvelzen2022revisiting, fleck2010reflecting}. 
% These systems could support the development of writing skills~\cite{magooda2022improving, neshaei2025mindmate}, promote self-awareness and well-being~\cite{kim2024mindfuldiary, kocielnik2018reflection}, and help train professional skills~\cite{arakawa2020inward, hoque2013mach, zhou2021virtual}.
% Furthermore, researchers have begun to integrate LLMs into these tools, leveraging their ability to generate personalized feedback and support conversational interactions that foster richer reflective experiences~\cite{arakawa2023catalyst, xu2024jamplate}.
% For example, \textit{Friction}~\cite{zhang2025friction} explored how LLMs can provide immediate, structured feedback in writing tasks, guiding authors toward instant and continuous reflection and experimentation. This work inspires us to explore the use of LLMs to provide teachers with instant and continuous opportunities for reflection.
\zhh{Reflective learning---the process of internally examining an experience to gain deeper understanding and new perspectives~\cite{boyd1983reflective}---has motivated HCI research on interactive systems that promote reflection~\cite{bentvelzen2022revisiting, fleck2010reflecting}, spanning writing~\cite{magooda2022improving, neshaei2025mindmate}, well-being~\cite{kim2024mindfuldiary, kocielnik2018reflection}, and professional skills~\cite{arakawa2020inward, hoque2013mach, zhou2021virtual}. 
For example, \citet{rich2009video} introduced a video annotation tool to assist teachers in reflecting on their classroom teaching and proved its effectiveness. 
Researchers have begun to leverage large language models to provide
% integrating LLMs into such tools for 
personalized feedback and richer reflective interactions~\cite{arakawa2023catalyst, xu2024jamplate}. For example, \textit{Friction}~\cite{zhang2025friction} provides LLM-driven immediate feedback in writing tasks, inspiring us to explore similar instant reflection support for teachers.
Additionally, \textit{TutorUp}~\cite{pan2025tutorup} leverages LLM-generated content to help tutors reflect on their practice to improve their ability to address engagement challenges within text-based simulated student environments.
However, previous reflective learning support tools focus on either pre-scripted or text-based settings, and they seldom look into learners' needs during VBL of instructional strategies. 
We conduct interviews with both early-stage and experienced teachers to inform the design of reflective support in \name{}'s watching, practicing, and evaluation stages.
}

% In the field of educator training, \citet{rich2009video} introduced a video annotation tool to assist teachers in reflecting on their classroom teaching and proved its effectiveness, related to the concept of VBL.
% Additionally, \textit{TutorUp}~\cite{pan2025tutorup} leverages LLM-generated content to help tutors reflect on their practice to improve their ability to address engagement challenges within text-based simulated student environments.
% However, there is still a lack of automated VBL tools that provide feedback to support teachers' professional development.
% Motivated by the potential of LLMs to provide timely support and foster reflective learning, 
% we apply them to VBL environments to enhance teachers' professional development.

\section{Formative Study}
\zhh{
% Prior work \fhx{[refs]} has shown that novice teachers often struggle to identify which videos are valuable for learning among a large number of videos when engaging in VBL for PD. 
% Additionally, they frequently lack the skills needed to engage in deep reflection during the process. 
% To gain a deeper understanding of the specific challenges teachers face and to derive precise design requirements for our system, we conducted a formative study involving nine teachers.
To inform the design of \name{}, we conducted a formative study with nine participants to understand their practices, challenges, and requirements in video-based learning of instructional strategies. 
% To understand teachers' practices, challenges, and requirements for a supporting tool in video-based learning (VBL) of teaching strategies, we conducted a formative study with nine participants, including six novice teachers (FN1-6) and three experienced teachers (FE1-3). The detailed information of the participants is shown in~\autoref{tab:formative} in appendix.
% Novices were either pre-service teachers or teachers with less than three years of experience, while experienced teachers had been organizers or mentors in programs of teachers' professional development for multiple times. 
% We used a snowballing approach to recruit these participants, starting from teachers in a local elementary school and students in a local normal university. 
}

\zhh{
\subsection{Participants and Procedure}
% \subsection{Participants and Procedure}
% As shown in Figure 1, the participants include five novice teachers (P1-P5) and four experienced teachers (E1-E4). 
% All experienced teachers have participated in teacher PD programs as organizers or mentors multiple times.
We used a snowballing approach to recruit nine participants (~\autoref{tab:formative} in Appendix), starting from a local normal university and an elementary school. 
Four of them (FN1-4) are our targeted early-stage teachers who are under training in teaching chemistry, history, physics, or music. 
We also include three very experienced teachers (FE1-3) with over 20 years of teaching experience, as well as two relatively experienced teachers (FE4-5) with three years of teaching experience in Chinese or Mathematics in the elementary school. 
These participants with varying teaching experiences can inform \name{}'s designs from both learners' and trainers' perspectives. 

We conducted one-on-one semi-structured interviews with each participant, either online
%(P5, P6, P7, P9)
\zzh{(FN1-4)}
or offline
%(P1, P2, P3, P4, P8)
\zzh{(FE1-5)}, lasting between 45 and 60 minutes. 
% We conducted one-on-one semi-structured interviews with each participant, both online (P5, P6, P7, P9) and offline (P1, P2, P3, P4, P8), lasting between 45 and 60 minutes. 
We began with participants sharing their experiences with and viewpoints on VBL. %, including video sources, process, and objectives. 
% Then, the focus narrowed to the perceived effectiveness of VBL, gathering distinct viewpoints from both trainees (novice teachers) and trainers (experienced teachers).
Then, we presented a low-fidelity prototype of a VBL support tool in a slide deck to gather feedback and design requirements.
% Finally, we used a pre-prepared slide deck—serving as a low-fidelity prototype, to gather feedback on the interface and functional design requirements.
\fhx{The prototype consists of two main pages: one for playing videos, which provides some hints and basic functionalities, 
and the other that outlines a possible user assessment.
The slide deck used in the formative study is shown in \autoref{fig:proto1} and \autoref{fig:proto2} in Appendix.}
We applied Braun and Clarke's six-phase thematic analysis framework~\cite{guest2011applied} to analyze the interview data. 
% To minimize bias, we incorporated peer debriefing throughout the process. 
One researcher initially coded all the qualitative data, and another carefully reviewed the codes to ensure accuracy and completeness. 
% Through iterative discussions, the two authors reached a consensus and identified four themes, which are related to what novice teachers focus on in classroom videos (Finding 1), the challenges of traditional VBL as a PD approach (Finding 2), and suggestions for the prototype (Finding 3).
Through iterative discussions, the two authors reached a consensus and identified three themes about early-stage teachers' focuses during learning, challenges, and suggestions for the VBL support tool. % what novice teachers focus on in classroom videos, the challenges of traditional VBL for professional development, and suggestions for the VBL support tool.
}
% \begin{table}[htbp] % 使用 [htbp] 让 LaTeX 自动选择最佳位置
% \centering
% \begin{threeparttable}
% \caption{Experienced and novice teachers involved in the formative study} % 为表格添加一个标题
% \label{tab:formative}   % 用于交叉引用
% \begin{tabular}{ccccc}
% \toprule % booktabs 的顶部线
% \textbf{ID} & \textbf{Gender} & \textbf{Educational Career Length} & \textbf{Educational Level} & \textbf{Subjects} \\
% \midrule % booktabs 的中间线
% FE1 & Female & 38 years & Teacher Development Center\tnote{1} & -- \\
% FE2 & Female & 21 years & Elementary School & Chinese \\
% FE3 & Female & 25 years & Elementary School & Mathematics \\
% \midrule % 使用 midrule 代替 hline
% FN1 & Male & 3 years & Elementary School & Chinese \\
% FN2 & Male & 3 years & Elementary School & Mathematics \\
% FN3 & Male & 1 year (Intern) & Senior High School (Intended) & Chemistry \\
% FN4 & Female & 1 year (Intern) & Senior High School (Intended) & History \\
% FN5 & Male & 2 years (Intern) & Junior High School (Intended) & Physics \\
% FN6 & Male & 1 year (Intern) & Senior High School (Intended) & Music \\
% \bottomrule % booktabs 的底部线
% \end{tabular}
% \begin{tablenotes}
%     \item[1] Teacher Development Center is an institution under the local Education Bureau that is responsible for curriculum research, teacher training, and professional development.
% \end{tablenotes}
% \end{threeparttable}
% \end{table}

\subsection{Findings and Design Requirements}

\zhh{
\textbf{Finding 1: Early-stage teachers focus on learning instructional strategies within the context of specific subjects.}
All early-stage teachers reported focusing on learning from videos about what instructional strategies were used and how to use them within subject-specific contexts. 
% An interesting finding is that when teachers described these strategies, they often did so within subject-specific contexts. 
For example, 
%P9
\zzh{FN2} said, \textit{``I would pay attention to how to give feedback on students' pitch problems and how to explain abstract concepts such as harmony.''}  
%p3
% \zzh{FE5} recalled his experience as a novice, 
% \textit{``in a video, a student presented a correct but unexpected method to solve a math problem; I thought of ways to connect it to the next part that I was going to teach.''} 
This finding reveals that when teachers notice instructional strategies in the videos, their observations on the use of these strategies are highly contextualized to the teaching content.
}

\zhh{
\textbf{Finding 2: Developing teaching competence in traditional VBL can be challenging for early-stage teachers.}
All early-stage teachers expressed a desire to develop a deeper understanding of instructional strategies during the VBL process. %, rather than simply copying all of a teacher's behaviors into their own classrooms. 
However, there are two challenges that hinder the development of teaching competence. 
First, traditional videos tend to showcase only surface-level activities while concealing the internal cognitive processes, thus lacking the ``thinking cues'' that would explain the pedagogical reasoning behind those actions. 
\zzh{FE2}
and 
\zzh{FE3}
highlighted the risk that early-stage teachers who only imitate the activities of expert teachers, without possessing the necessary skills, may struggle to maintain classroom control.
% \zzh{FE3} provided an example: \textit{``(a novice teacher in training) imitated a famous teacher (the lesson), but during class, the students' behavior differed from what he had anticipated, and he didn't know how to manage the class... actually, it was because he lacked the ability to handle that particular class.''}
Second, six participants 
\zzh{(FE1, FE2, FE4, FE5, FN1, FN2)}
noted that the instructional strategies and pacing in the videos were tailored to students with specific prior knowledge and learning abilities, which often did not match their own classroom contexts---especially in award-winning open classes featuring highly cooperative students.
\zzh{FE4}, who already had three years of teaching experience,
noted, \textit{``Sometimes when watching how famous teachers teach in developed regions, I realized that the students are in very good condition. What if I face students with weak knowledge foundations?''}
% Overall, novice teachers struggle to develop teaching competence in VBL due to a cognitive gap in the videos and a contextual gap between the on-screen students and their own.
}

\zhh{
\textbf{Finding 3: \name{} should provide reflective support throughout the video-based learning process.}
% \textbf{Finding 3: The VBL support tool should facilitate instant reflection throughout watching and practicing stages.}
After reviewing the prototype slide deck we provided, both early-stage and experienced teachers recognized the potential of reflection and evaluation to enhance VBL outcomes.
% Specifically, participants 
%E4
\zzh{FE2}
and 
%P7
\zzh{FN1}
suggested that the system should prompt users to reflect right after watching a video and after they practice teaching on camera, rather than only providing feedback at the end. 
% Furthermore, 
%E1 and E8
% \zzh{FE1 and FE3}
% emphasized that reflection activities provided to users should be instant, as 
% %E1
% \zzh{FE1}
% saying, \textit{``Only if feedback is instant will (novice teachers) be interested in using it.''}
}

\zhh{
The focus on learning instructional strategies within subject-specific contexts (Finding 1) and need to uncover the pedagogical reasoning underlying teachers' actions point to our first design requirement: \textbf{DR1) \name{} should help users \haoxiang{accumulate observations of} subject-specific contextual teaching strategies \haoxiang{from} the videos.} 
This requirement aligns with pedagogical content knowledge that integrates pedagogy and disciplinary content to promote professional development~\cite{shulman1986those}. 
Besides, as early-stage teachers often lack teaching competence and would face classroom contexts different from the videos' (Finding 2), personalized practices and reflective feedback based on the video content (Finding 3) are needed: \textbf{DR2) \name{} should provide personalized training tasks and feedback to bridge the gap between VBL and real classroom teaching.} 
Lastly, as reflection is a key mechanism for transforming experience into learning~\cite{kolb2014experiential,boud2013reflection} and participants expect reflective support throughout the learning process (Finding 3), we have: \textbf{DR3) \name{} should encourage continuous and timely reflection to enhance early-stage teachers' understanding of instructional strategies throughout the VBL process.}
}

\zhh{
\subsection{Marzano's Nine Instructional Strategies}\label{sec:strategies}
\zhhred{As suggested by our highly experienced teachers (FE1-3),} 
\name{} aims at supporting the learning of nine instructional strategies, which are identified by \citet{marzano2001classroom} to be most effective among forty strategies in improving students' outcomes in the classroom. %~\autoref{tab:strategies}.
% \haoxiang{
These categories are especially suitable here because they describe observable, cross-subject instructional moves that can appear as video-identifiable classroom events and can later be rehearsed in practice~\cite{marzano2001handbook}.
The nine strategies are: 
(1) identifying similarities and differences; 
(2) summarizing and note taking; 
(3) reinforcing effort and providing recognition; 
(4) homework and practice; 
(5) nonlinguistic representations; 
(6) cooperative learning; 
(7) setting objectives and providing feedback; 
(8) generating and testing hypotheses; and 
(9) activating prior knowledge with cues, questions, and advance organizers. 
The detailed definitions and some examples of the nine strategies are shown in~\autoref{tab:strategies} in Appendix.
This same framework has also been extended to technology-mediated instruction, including work that maps the nine categories to classroom technologies, online / blended course design, and empirical studies of teacher uptake~\cite{pitler2012using,bolt2021developments,almekhlafi2020teachers}.
}
\section{\name{} Design and Implementation}
\label{sec:design}
% The design requirements derived from our formative study (DR1-DR3) center on identifying what and how teaching strategies are used in classroom videos, such that we can help users analyze these strategies (DR1), provide personalized training tasks (DR2), and encourage continuous reflection (DR3).
% Accordingly, 
\zhh{
In this section, we first describe a user scenario to walk through \name{}'s interface design, followed by its implementation. 
}

% \haoxiang{This section first presents our computational pipeline that automatically detects nine instructional strategies and extracts relevant video clips.
% Built upon these video analysis technology, we introduce \name{}'s system design, which supports teachers' professional development through a three-stage learning process: watching, practicing, and evaluating.
% }

\zhh{
\subsection{Interface and Interaction Design}
\label{sec:system-interface}
}

% \haoxiang{
% \subsubsection{User Workflow and Stage Design.}
% \label{sec:usage-scenario}
% \label{sec:stage-design}
% }

\begin{figure*}[htbp]
\centering
\includegraphics[width=0.88\linewidth]{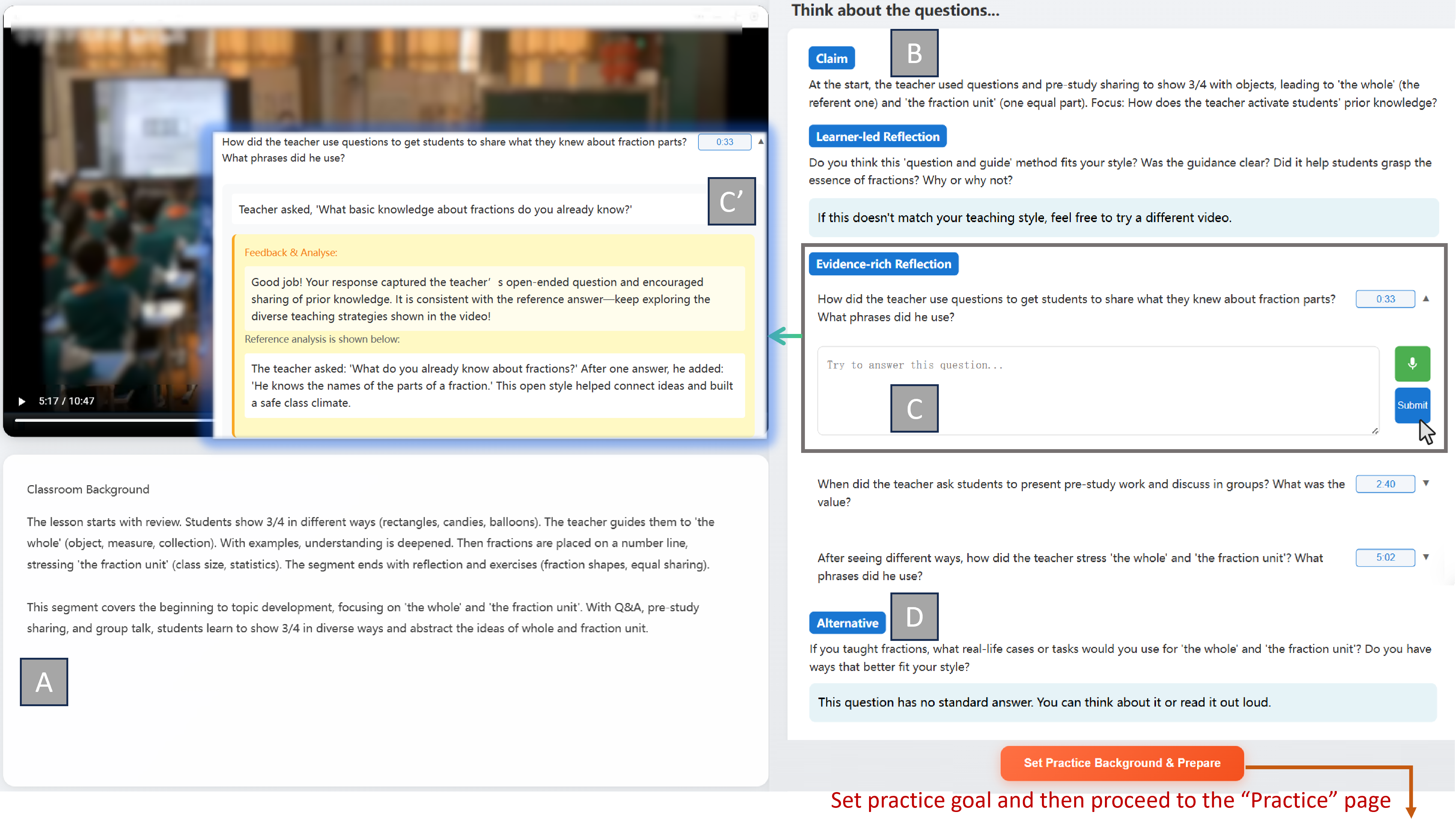} 
\caption{\haoxiang{\name{}'s interface in the watching stage.
The interface is originally in Chinese and translated by GPT-4.1.} 
}
\Description{TeachUp’s interface in the watching stage. The interface is originally in Chinese and translated by GPT-4.1.}
\label{fig:watch}
\end{figure*}

\begin{figure*}[htbp]
\centering
\includegraphics[width=0.88\linewidth]{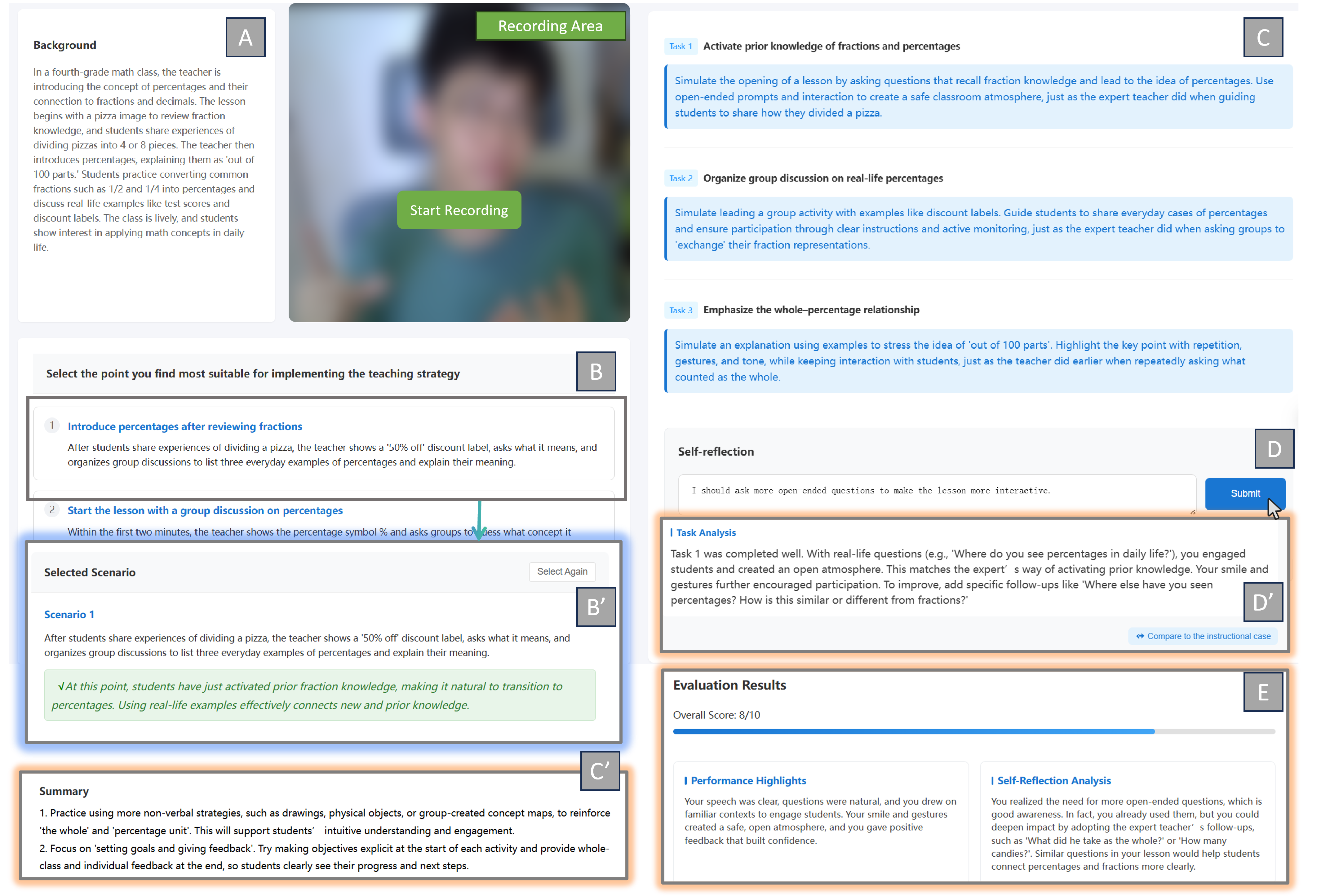} 
\caption{
\haoxiang{
\name{}'s interface in the practicing stage (A, B, C, D) and evaluating stage (A, B', C', D', E).
The interface is originally in Chinese and translated by GPT-4.1.} 
}
\Description{TeachUp’s interface in the practicing stage (A, B, C, D) and evaluating stage (A, B’, C’, D’, E). The interface is originally in Chinese and translated by GPT-4.1.}
\label{fig:practice}
\end{figure*}
% [Merged from the original 4.2.1 (User Workflow: A Usage Scenario) and 4.2.2 (Design of the Three Learning Stages) to reduce redundancy.]

% \subsubsection{User Workflow: A Usage Scenario.}
% \label{sec:usage-scenario}
\zhh{
Consider Harry, an early-stage teacher who just graduated from a normal university and will teach fourth-grade mathematics for the first time next semester. 
He turns to \name{}, hoping to learn how to effectively \textit{activate students' prior knowledge} from others' classroom videos. 
After selecting a 10-minute video clip which is tagged with this instructional strategy and about ``The Meaning of Fractions'' in \name{}'s database, Harry engages in the following three learning stages. 
% We walkthrough \name{}'s interface and interaction design via a user scenario of Mr.\ Smith, a novice teacher who will be teaching fourth-grade mathematics for the first time next semester.
% He hopes to improve his pedagogical content knowledge, especially in effectively activating students' prior knowledge.
% Therefore, he begins to use \name{}.
}

\zhh{
\textbf{Watching a video clip related to targeted strategy.} 
As shown in \autoref{fig:watch}, Harry can review the classroom background (A), a well-contextualized description generated by analyzing the selected clip in relation to its full lesson video (DR1). 
At any time of the watching stage, he can check a series of reflective hints (DR3), based on a revised version of the Lesson Analysis Protocol (LAP)~\cite{taylor2017effect}. 
These include a claim identifying the target strategy and a learner-led reflection on its suitability (B), evidence-rich reflection questions with clickable timestamps that jump the video to specific moments (C), and an alternative prompt encouraging him to articulate his own lesson design ideas (D). 
Harry can respond to the reflection question and get feedback (C'). 
He can click ``Set Practice Background \& Prepare'' to proceed to the ``Practice'' stage. 
% After each response, he receives instant system feedback (C'). Through this stage, Mr.\ Smith fully comprehends the expert's strategy.
}

% \haoxiang{In the \textbf{watching} stage (\autoref{fig:watch}), he reviews the classroom background (\autoref{fig:watch} A), a well-contextualized description generated by analyzing the selected clip in relation to its full lesson video (DR1).
% He then works through a series of reflective hints (B--D), based on a revised version of the Lesson Analysis Protocol (LAP)~\cite{taylor2017effect} (DR3):} a claim identifying the target strategy, a learner-led reflection on its suitability, evidence-rich reflection questions with clickable timestamps that jump the video to specific moments, and an alternative prompt encouraging him to articulate his own lesson design ideas. After each response, he receives instant system feedback (C'). Through this stage, Mr.\ Smith fully comprehends the expert's strategy.

% \textbf{Practicing stage: microteaching practice} \haoxiang{(\autoref{fig:practice} A, B, C, D)}. 
% \haoxiang{To satisfy DR2, we adapted the Microteaching Cycle, which emphasizes the iterative process of \textit{plan -> teach -> feedback}~\cite{allen1972microteaching}}.
% \haoxiang{Microteaching is a teacher training approach in which instructors practice short, focused lessons in a controlled setting to refine specific teaching skills~\cite{remesh2013microteaching}. Users can customize the classroom background (\autoref{fig:practice} A, B), study the tasks (C) provided by \name{}, and then conduct microteaching recordings in front of the camera.}

\zhh{
\textbf{Practicing the learned strategy}. 
\name{} adapts the Microteaching Cycle (\textit{plan $\to$ teach $\to$ feedback})~\cite{allen1972microteaching,remesh2013microteaching} to support personalized training (DR2). 
Specifically, Harry sets a practice goal of teaching ``the meaning of percentages'' and receives a generated related teaching scenario  (\autoref{fig:practice} A), along with three possible starting points for implementing the teaching strategy (B). 
He selects ``Introduce percentage after reviewing fractions'' (B') and makes plans for three related tasks (C) --- two emphasizing imitation of the expert's strategy and one focusing on language, gestures, and tone. 
Once ready, Harry conducts a four-minute microteaching session. After that, he records a self-reflection and submits the practice (D).
}
% As shown in \autoref{fig:practice}A--D, 
% }

% \haoxiang{Mr.\ Smith then moves on to the \textbf{practicing} stage (\autoref{fig:practice} A--D), which adapts the Microteaching Cycle (\textit{plan $\to$ teach $\to$ feedback})~\cite{allen1972microteaching,remesh2013microteaching} to satisfy DR2.}
% He sets his practice goal to teach ``the meaning of percentages'' by clicking the button ``Set Practice Background \& Prepare''.
% % (\autoref{fig:watch} E). 
% The system creates a teaching scenario (\autoref{fig:practice} A) with three possible starting points (B); he selects the most suitable one and receives confirmation (B'). Guided by three tasks (C)---two emphasizing imitation of the expert's strategy and one focusing on language, gestures, and tone---he conducts a four-minute microteaching session and then records a self-reflection (D).

% \textbf{Evaluating stage: feedback connecting instructional case and practice} \haoxiang{(\autoref{fig:practice} A, B', C', D', E)}. 
% After analysis, \name{} provides overall scores, evaluations, and suggestions on teachers' performances (DR3, \haoxiang{\autoref{fig:practice} D'}).
% This feedback covers teachers' verbal content, tone, gestures, and facial expressions, and is grounded in their practicing videos and the instructional case from the watching stage, allowing them to develop more concrete, in-context reflections through comparison with expert practice.

\zhh{
\textbf{Evaluating performance of practiced strategy.}
After submission, Harry will receive a personalized report on his task performance (DR2, DR3). 
The report covers verbal content, tone, gestures, and facial expressions (\autoref{fig:practice} C'), and is grounded in his practicing video and the watched clip.
It highlights Harry's strengths based on multimodal analysis, expands on his self-identified areas for improvement with suggestions grounded in the instructional case, and provides task-by-task feedback (D') with a ``Compare to the instructional case'' button for side-by-side review. 
Finally, Harry can check a summary that recommends specific categories of strategies to study next. 
Harry now has learned how to effectively \textit{activate students' prior knowledge} from the classroom video. 
}

% \haoxiang{About three minutes later, the system delivers a personalized \textbf{evaluating} report (DR3, \autoref{fig:practice} D'). This feedback covers verbal content, tone, gestures, and facial expressions, and is grounded in his practicing video and the instructional case from the watching stage. Specifically, it} highlights his strengths based on multimodal analysis, expands on his self-identified areas for improvement with suggestions grounded in the instructional case, and provides task-by-task feedback (C') with a ``Compare to the instructional case'' button for side-by-side review. Finally, a summary (E) recommends specific categories of teaching strategies for him to study next.

% Through this complete three-stage process of watching, practicing, and evaluating, Mr. Smith not only gains a deep understanding of the expert teacher's strategy but also clearly recognizes his own strengths and weaknesses through the practice and personalized feedback.

% \haoxiang{
% \subsection{Detecting Instructional Strategies in Classroom Videos}
% \label{sec:pipeline}
% }

\zhh{
\subsection{System Implementation}
}
\subsubsection{Detecting Instructional Strategies in Classroom Videos}
\label{sec:pipeline}
% \zhh{
% To prepare a database of video clips tagged with instructional strategies (\autoref{sec:strategies}) in \name{}, we developed a strategy detection pipeline.% based on video transcripts. 
% }
\fan{To enable the system to provide relevant examples, a strategy-detection pipeline that prepares a database of candidate video clips for the instructional strategies in \autoref{sec:strategies} is needed.}

\zhh{
\textbf{Benchmark. }
We randomly selected a grade (Grade 1 to Grade 12), a subject, and then one textbook unit as the candidate topic. We searched the topic keywords on \textit{Bilibili} and collected three videos per topic that satisfied the criteria of ``40--50 minutes in length'' and complete classroom recordings under the default search setting. In total, three units were selected, yielding nine videos.
Two authors 
% \zhhred{[@Haoxiang, any training or guidance from experts? No, how to explain?]}
independently annotated the videos after agreeing on the classification criteria. To assess annotation reliability, segments with identical labels and temporal $IoU \ge 0.5$ were considered the same instructional strategy instance. 
The two annotators agreed on 85 instructional events; one annotator additionally marked 1 event and the other 4 events missed by their counterpart, corresponding to an agreement rate of 94.4\%. 
The remaining disagreements (five segments) were resolved through discussion.% to establish the final benchmark.  
Because the task involves detecting multiple temporal segments in long videos rather than assigning a single label per instance, traditional inter-rater statistics such as Cohen’s $\kappa$ or ICC are not applicable.
The final benchmark contains 89 strategy instances in nine instructional strategy categories, with 6--15 segments per category. 
We released this benchmark as supplementary material.
}
% \fan{To enable the system to provide relevant examples, we developed a strategy-detection pipeline that prepares a database of candidate video clips for the instructional strategies in \autoref{sec:strategies}.
% }

% \subsubsection{\haoxiang{Pipeline Architecture}}
\begin{figure}[]
\centering
\includegraphics[width=1\linewidth]{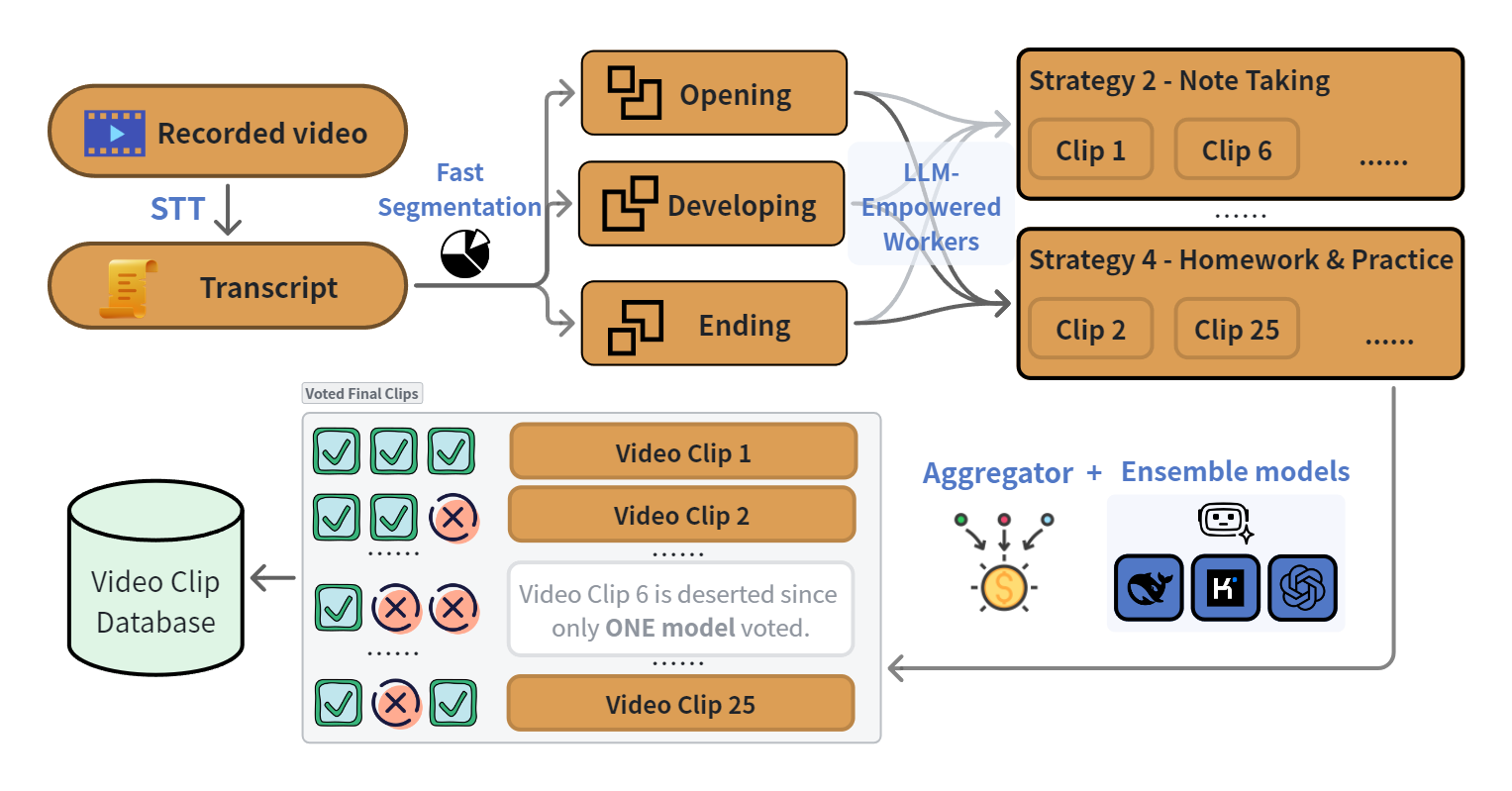} 
\caption{The computational pipeline of detecting instructional strategies in classroom videos.}
\Description{The computational pipeline of detecting instructional strategies in classroom videos.}
\label{fig:pipeline}
\end{figure}

\zhh{
\textbf{Pipeline}. 
% Two authors iteratively annotated 89 instructional strategies within nine videos (40–50 minutes each), which were used to build a benchmark (\autoref{sec:benchmark}).
On the benchmark, we explored multiple modalities for instructional strategy detection, including vision (\eg large vision-language models, pose detection, emotion recognition), audio (\eg Qwen-2-Audio), and text, as well as different workflow architectures.
% However, because our target setting involves large-scale video processing, fully multimodal solutions are less practical in terms of efficiency and deployment cost. 
\zhhred{However, including more modalities does not improve performance of our pipeline, potentially because of the small size of benchmark.}
Balancing scalability, efficiency, and accuracy, we finally adopted a pipeline that primarily relies on transcribed spoken text as input. 
% , consisting of three main blocks
As shown in \autoref{fig:pipeline}, 
the pipeline inputs the sentence-level transcript (by Tingwu Big Model~\footnote{\url{https://tingwu.aliyun.com}}) of a classroom teaching video and outputs a set of video clips ($s_i$), each with a strategy category ($\text{cat}_i$), a confidence score ($\text{conf}_i$), and precise timestamps ($t_{s_i}, t_{e_i}$). 
We first conduct a fast segmentation, as the nine teaching strategies are typically associated with different lesson stages, \eg ``cues, questions, and advance organizers'' often occur at the beginning, while ``homework and practice'' typically appears in the middle to late stages. 
Specifically, we prompt an LLM to coarsely divide the lesson into three stages: the \textit{opening} spans from the start of class to the last review of prior knowledge; the \textit{developing} covers the introduction and teaching of new content; and the \textit{ending} begins with the first in-class exercise. 
The three stages overlap partially to reduce missed detections. 
Then, we designed 
% five \zhhred{[Why five?]}
parallelized LLM workers that collectively detect the nine instructional strategies, with each worker responsible for detecting events in its assigned lesson stage. Each worker outputs strategy segments in the form $(t_{s_i}, t_{e_i}, \text{cat}_i)$. 
Lastly, we designed a rule-based aggregator. For any two segments $S_i=(t_{s_i}, t_{e_i})$ and $S_j=(t_{s_j}, t_{e_j})$ with the same category $\text{cat}_k$, if $t_{s_i} \le t_{e_j} \land t_{s_j} \le t_{e_i}$, they are merged into $S_{\text{new}}=(\min(t_{s_i}, t_{s_j}), \max(t_{e_i}, t_{e_j}))$. Merging is repeated until convergence, producing the final output $[(s_1, \text{cat}_1, \text{conf}_1), (s_2, \text{cat}_2, \text{conf}_2), \dots]$.
}

\zhh{
% A \textbf{prompt-based few-shot learning approach} is used 
To improve detection precision, we first applied a zero-shot prompt on eight carefully selected high-quality videos and analyzed the models' typical errors (\eg for cooperative learning, the model only detected ``how to organize group discussions'' but missed ``how to evaluate the outcomes''). We then incorporated transcript segments as positive few-shot examples and frequently misclassified segments as negative \textbf{few-shot} examples. 
For each event, we designed two to four few-shot examples.
% We developed a prompt-based few-shot learning approach through an iterative process to achieve precise detection.
% First, we applied a zero-shot prompt to perform multiple inferences on eight carefully selected high-quality videos and analyzed the model's typical errors (\eg for cooperative learning, the model often detected ``how to organize group discussions'' but missed ``how to evaluate the outcomes'').
% Then, we gathered exemplary transcript segments from the inferred videos and incorporated them into the prompt as positive few-shot examples, while including frequently misclassified transcript segments as negative few-shot examples.
% For each event, we designed between two and four different few-shot learning examples.
To improve reliability, we applied \textbf{ensemble} with three LLMs---DeepSeek-v3~\cite{deepseekai2024deepseekv3technicalreport}, Kimi-k2, and GPT-4.1---in parallel to determine the final video clips. % through greedy clustering. 
% This ensemble improves reliability through greedy clustering. \textbf{Event Clustering:} 
Specifically, for each instructional event type, predictions from all models are clustered. The first prediction initializes a cluster, and each subsequent event is merged if its temporal overlap ratio exceeds 0.5; otherwise, it starts a new cluster. 
% \textbf{Voting:} 
A cluster is accepted only if it contains predictions from at least two models. % ($\texttt{VOTING\_THRESHOLD}=2$). 
% \textbf{Boundary and Confidence Fusion:} 
For each accepted cluster, the final boundary is defined by the earliest start and latest end times, and the confidence score is the average of member scores.}
% We apply an ensemble method that integrates three LLMs, DeepSeek-v3~\cite{deepseekai2024deepseekv3technicalreport}, Kimi-k2, and GPT-4.1, in parallel to decide the final video clips.
% This ensemble approach enhances reliability through a greedy clustering algorithm.
% \begin{itemize}
%     \item \textbf{Event Clustering:} For each instructional event type, predictions from all models are clustered. The first prediction initializes a cluster, and each subsequent event is merged if its temporal overlap ratio exceeds 0.5; otherwise, it starts a new cluster.
%     \item \textbf{Voting.} A cluster is accepted only if it contains predictions from at least two models ($\texttt{VOTING\_THRESHOLD}=2$).
%     \item \textbf{Boundary and Confidence Fusion.} For each accepted cluster, the final boundary is defined by the earliest start and latest end times, and the confidence score is the average of member scores.
% \end{itemize}
% The ensemble method filters out errors from individual models, resulting in more robust outcomes.

% \subsubsection{\haoxiang{Technical Evaluation}}
\zhh{
\textbf{Technical evaluation.} 
We evaluate the proposed pipeline using IoU, Precision, Recall, and F1-Score. For a predicted segment $S_{pred}$ and a ground-truth segment $S_{gt}$,}
% To evaluate the effectiveness of the proposed pipeline, we conducted a technical evaluation that involved selecting appropriate metrics, constructing a benchmark, and analyzing the results. To measure the alignment between the model's predicted segments and the manually annotated ground truth, we used the Intersection over Union (IoU) threshold as the criterion for successful classification. For a predicted time segment $S_{pred}$ and a ground-truth segment $S_{gt}$, the IoU is calculated as:
% $$ IoU = \frac{|S_{pred} \cap S_{gt}|}{|S_{pred} \cup S_{gt}|} $$
\zhh{We prioritized precision because the downstream task requires reliable instructional segments for learners, even at the cost of lower recall.
We compared the ensemble with individual models on the benchmark across IoU thresholds from 0.1 to 0.8. The ensemble consistently achieves higher precision than individual models, while performance for all models gradually decreases as the IoU threshold becomes stricter.}
\zhh{At $IoU=0.5$---a common threshold in object detection~\cite{everingham2010pascal}---the ensemble achieves a precision of 0.634 of detecting nine strategies. 
% , outperforming all individual models, suggesting that it effectively filters incorrectly detected segments. 
In contrast, a zero-shot baseline using a single LLM prompted only with strategy definitions performed poorly (precision = 0.095, recall = 0.191, F1 = 0.127, at $IoU=0.5$).
% Overall, \zhhred{this pipeline achieves acceptable performance that supports \name{}'s usage in our later interviews with teachers, while in our within-subject study we control the used video clips with ground-truth strategies.
% We will release the benchmark and pipeline upon which future work can improve. 
\fan{Overall, these results provide an initial feasibility check of the pipeline as an enabling component for \name{}.
% Because the within-subjects study used clips with ground-truth strategy labels, it did not evaluate how pipeline errors might affect learning.
}
} 

\begin{figure}[]
\centering
\includegraphics[width=1\linewidth]{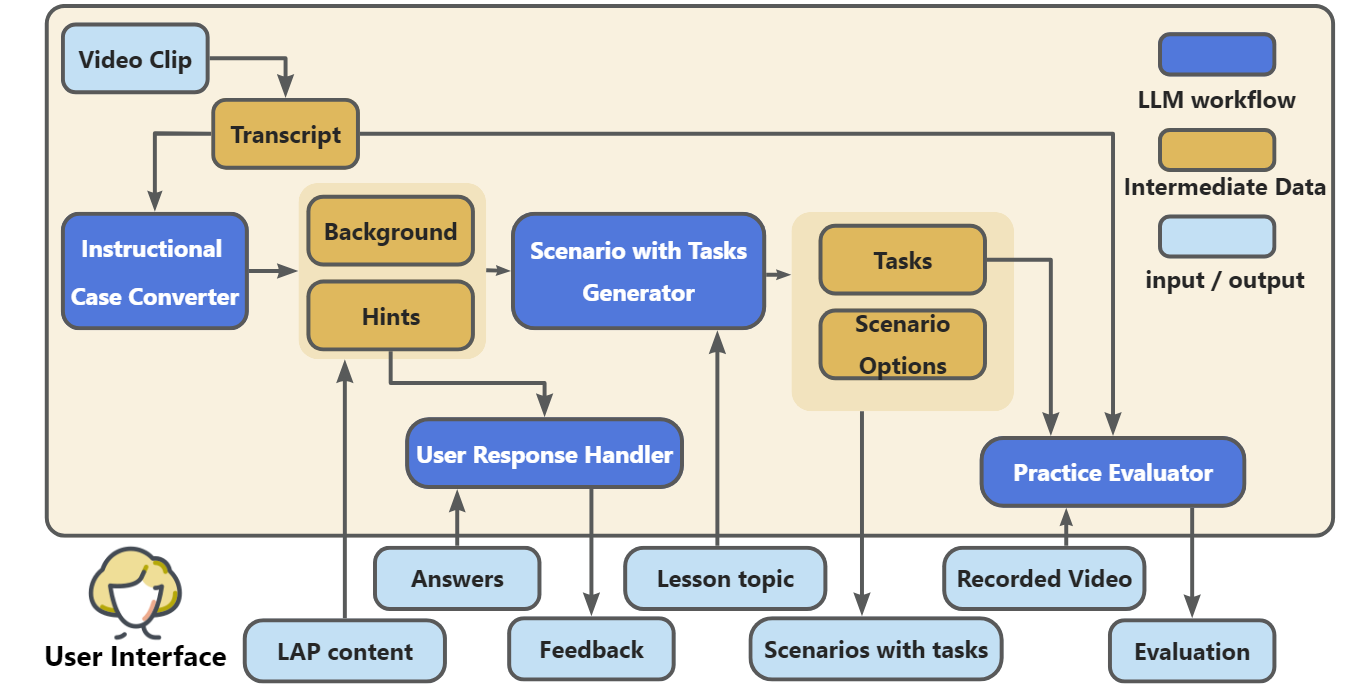} 
\caption{\haoxiang{The architecture of \name{}. The Video Clip is the output of the pipeline introduced in \autoref{sec:pipeline}.}}
\Description{The architecture of TeachUp. The Video Clip is the output of the pipeline introduced in Section 4.2.1.}
\label{fig:sysarch}
\end{figure}

\subsubsection{\zhh{Supporting User Interactions}}
\zhh{
As shown in \autoref{fig:sysarch}, we developed four modules that enable user interaction with \name{}. 
These modules are driven by an LLM (GPT-4.1) and a multimodal LLM (Qwen-2.5-Omni~\cite{Qwen2.5-Omni}), connected through a modular workflow that balances context richness and response latency.
}
% \haoxiang{
% The system architecture of \name{} consists of five components: an instructional strategy detection pipeline (\haoxiang{\autoref{sec:pipeline}}), an instructional case converter, a user response handler, a scenario-with-tasks generator, and a practice evaluator.}

% \haoxiang{
% The workflow proceeds as follows. First, the detection pipeline extracts video clips containing instructional strategies. These clips are converted into instructional cases that provide background information and hints for the watching stage. Based on the user's reflections, the system generates practice scenarios and tasks. The user’s responses are processed for feedback, and the user's microteaching video is evaluated using a multimodal model. 
% The system is driven by an LLM (GPT-4.1) and a multimodal LLM (Qwen-2.5-Omni~\cite{Qwen2.5-Omni}), \haoxiang{connected through a modular workflow that balances context richness and response latency}.}

% \textbf{{\haoxiang{Instructional Case Converter.}}}
\zhh{
With the user-selected video clip, 
the \textit{Instructional Case Converter} transforms it into instructional cases, including background explanations and hints. 
To minimize user waiting time, we pre-process all video clips offline before releasing \name{} to users. 
% we adopt a pre-generation strategy: all clips are processed offline before system release.
% Since this step involves long LLM chains and large contexts, pre-computation ensures instant delivery during the watching stage.
}
% \textbf{{\haoxiang{User Response Handler.}}}
\zhh{
Once users respond to a reflection question in the watching stage, the \textit{User Response Handler} provides immediate feedback using the user’s answer, the claim, reflection questions, and reference analysis, excluding video-related information such as transcripts or background.
Keeping the prompt short reduces LLM latency and enables real-time feedback.
}
% \textbf{{\haoxiang{Scenario with Tasks Generator.}}}
\zhh{
The \textit{Scenario with Tasks Generator }generates practice scenarios and associated tasks by prompting an LLM to remain neutral and objective, avoiding evaluative cues (\eg ``however'', ``although...but...'') that might reveal the correct answer.
This prompt includes the background and hints from the watching stage, so that the generated tasks reflect the instructional strategies observed in the video.}
\zhh{
The \textit{Practice Evaluator} uses a multimodal LLM to evaluate users' microteaching practices. 
Because our used model accepts at most 40 seconds of video, the recording is resized and divided into 30-second windows with a 5-second overlap.
For each segment, the model outputs a JSON containing a transcript, audio–visual observations, and evaluation suggestions.
Segment-level results are concatenated and summarized into a final assessment.
To reduce inference time, multimodal evaluation runs in parallel workers, enabling a full microteaching video to be processed in under five minutes.
}

\section{User Study}
% To evaluate \name{}, we conducted two stage of user study to answer the following research questions (RQs):
% \begin{itemize}
%     \item \textbf{RQ1:} What are users’ perceptions of \name{} in terms of usability and usefulness?
%     \item \textbf{RQ2:} How effective does \name{} support teachers to learn teaching strategies from classroom teaching videos?
% \end{itemize}

\zhh{
We conducted a within-subjects study with 16 early-stage teachers to evaluate \name{}, guided by two research questions: 
% To evaluate \name{}, 
% we conducted a two-stage user study to answer the following research questions (RQs): 
\textbf{RQ1:} How effectively does \name{} support teachers in learning instructional strategies from classroom videos?
\textbf{RQ2:} What are users’ perceptions of \name{} in terms of usability and usefulness? 
% \textbf{RQ2:} How effectively does \name{} support teachers in learning teaching strategies from classroom videos?
}

% First, we conducted a within-subjects controlled study with 16 pre-service teachers using videos \pzh{of pre-set subjects}, examining the RQs both qualitatively and quantitatively.
% To further assess the usefulness of our proposed video processing pipeline and to gain insights from the perspective of in-service novice and expert teachers, we performed an interview study, in which participants were allowed to freely explore the system \pzh{with videos about their teaching subjects.}
\zhh{
\subsection{Participants}
% \subsection{Within-subjects Controlled Study}
% \subsubsection{Participants and Tasks Assignment.}
We recruited 16 participants (P1-P16, 8 males and 8 females, age: Mean = 22.19, SD = 2.01) through word-of-mouth. % and questionnaires.
\fan{All participants were pre-service teachers recruited from normal universities; P9--P16 additionally reported teaching experience, including internship experience (see \autoref{tab:exp_design_detailed} in the Appendix for detailed backgrounds).}
% All participants (detailed background in \autoref{tab:exp_design_detailed} in appendix) are either under training in universities \zhhred{(N=8)} or teachers with no more than one year of teaching experience including intern experience \zhhred{(N=8)}.
% The participants' backgrounds varied, including education majors (n=5), a certified teacher (n=1), participants with over four months of teaching experience (n=7), and frequent users of video-based learning learning to enhance their instructional skills (n=3).
% We also collected data on their experience with video-based learning and LLMs.
Five participants reported conducting video-based learning for their professional development weekly, four monthly, two yearly, three rarely, and two never. 
Twelve participants are daily users of large language models, and the remaining four are weekly users.
% For video-based learning, the reported frequency ranged from rarely to weekly; all participants reported using LLMs more often than weekly. 
% The details of the participants are shown in~\autoref{tab:exp_design_detailed} in appendix.
}

\subsection{Experimental Setup}
% The experiment was in a within-subjects design where each participant completed two teaching strategy learning tasks: one with \name{} and one with a baseline system.

% \begin{itemize}
%     \item \textbf{\name{} Condition.} Participants used \name{} with the ``watching - practice - evaluation'' stage within one instructional case. Additionally, the topic of the practice is fixed as we described below.
%     \item \textbf{Baseline Condition.} 
%     Have a similar user interface of \name{} but without AI-generated hints (the right panel of \autoref{fig:watch}), practice tasks (\autoref{fig:practice}  C), and evaluation (\autoref{fig:practice} C', D', E'). 
%     To ensure a fair comparison on the practice content and to save participants' time, we also provide the description of the most suitable scenario (\autoref{fig:practice} B') with participants.
% \end{itemize}

\zhh{
% \textbf{\name{} Condition.} 
\subsubsection{Conditions}
In the \textbf{\name{} condition}, participants used \name{} following the ``watching–practice–evaluation'' stages within one instructional case. 
% The topic of the practice task was fixed as described below.
In the \textbf{baseline condition}, participants used a baseline system with a user interface similar to \name{} but without AI-generated hints (\autoref{fig:watch} B, C, D), practice tasks (\autoref{fig:practice} C), and evaluation (\autoref{fig:practice} C', D', E).
To ensure a fair comparison and reduce participant workload, we provided the description of the most suitable scenario (\autoref{fig:practice} B') in the baseline condition.
}

% \haoxiang{\textbf{Baseline Condition.} The baseline system had a user interface similar to \name{} but without AI-generated hints (right panel of \autoref{fig:watch}), practice tasks (\autoref{fig:practice} C), and evaluation (\autoref{fig:practice} C', D', E'). To ensure a fair comparison and reduce participant workload, we provided the description of the most suitable scenario (\autoref{fig:practice} B').}

\zhh{
\subsubsection{Task}
We controlled the subjects and instructional strategies to compare participants' learning outcome and experience in two conditions. 
Specifically, we carefully selected four easy-to-understand content knowledge topics about Chinese Language and Mathematics in elementary schools. 
The Mathematics Task focused on the strategy of \textit{Cues, Questions, and Advance Organizers}, with \textit{Meaning of Fractions} as the VBL content and \textit{Meaning of Percentages} for the test after the learning session. 
The Chinese Language Task focused on \textit{Cooperative Learning}, with \textit{The Feet of the Boston Ivy} as the VBL content and \textit{The Cricket's Dwelling} for the test. 
The orders of participants' tasks and used systems were counterbalanced. 
% \zhhred{We selected one video clip around 15 minutes for each learning task\footnote{[The URLs of the videos are provided in the supplementary materials.}.
}

\subsubsection{Procedure.}
% Each experimental session was conducted either offline or online and lasted 80-90 minutes. One day before the session, participants received preparatory materials and were asked to spend at least 10 minutes reviewing them, which included
% \begin{itemize}
%     \item Brief descriptions of the two teaching strategies.
%     \item PDFs of textbooks and accompanying lesson plans with key content highlighted by one of the authors.
% \end{itemize}
\zhh{Each experimental session was conducted either offline or online and lasted 80--90 minutes. One day before the session, participants received
% preparatory materials---
brief descriptions of the two instructional strategies and annotated PDFs of the relevant textbooks and lesson plans.
They were asked to spend at least 10 minutes reviewing them.}
% On the day of the experiment, we first introduced the overall goal.
% Participants were prompted to take on the role of a novice teacher preparing for an upcoming class. Their main task was to learn and practice a key teaching strategy using the provided system for that condition. All participants received 100 RMB as compensation, and the top three performers received an additional 50 RMB bonus.
% Each condition began with a 10-minute introductory phase. 
% In this phase, a researcher explained the concept of teaching strategies to the participants and demonstrated the corresponding system. 
% After the introduction, participants proceeded to the system usage phase. They were asked to spend 12 minutes watching instructional examples and taking notes freely, followed by orally answering the question: \textit{``What did you learn from the video about the instructional strategies?''} This was followed by a 10-minute simulated teaching practice. In the \name{} condition, participants were given an additional three minutes to review a system-generated evaluation report on their practice. This step was included as it represents an integral part of the \name{}'s VBL cycle.
% At the end of the system usage phase, participants filled out a questionnaire. 
% Finally, in each condition, there was a performance assessment phase: participants were given a new scenario and, after three minutes of preparation, recorded a 3-5 minute teaching demonstration.
\zhh{On the day of the experiment, participants were prompted to prepare themselves for a classroom lecture. 
% introduced to the goal and prompted to take on the role of a novice teacher preparing for class. 
As an incentive, participants with top-3 performances in the test would receive a 50 RMB bonus in addition to 100 RMB compensation.
% All participants received 100 RMB as compensation, and the top three performers received an additional 50 RMB bonus. 
Each condition began with a 10-minute introduction in which a researcher explained the teaching strategies and demonstrated the system. Participants then spent 12 minutes watching instructional examples, followed by orally answering \textit{``What did you learn from the video about the instructional strategies?''} and a 10-minute simulated teaching practice. 
In the \name{} condition, participants additionally spent three minutes reviewing the system-generated evaluation report. At the end of each condition, participants completed a questionnaire and a teaching test, in which they were given a new scenario, three minutes to prepare, and recorded a 3--5 minute teaching demonstration.
There was a 10-minute break between the two conditions. The entire procedure concluded with a 10-minute interview, asking about participants' perceptions of the system,  usability of AI-generated content, opinions on the hints and feedback provided at each stage of the system for improving their learning outcomes.
Throughout the entire process, we recorded both the screen and audio for subsequent analysis.
}

\zhh{
\subsubsection{Measurements}
For \textbf{RQ1}, inspired by~\citet{wu2025comviewer}, we examined participants’ verbal reflection of what they learned from the video about the instructional strategies after the watching stage. 
We coded the number of \textit{valid inspiration points} %, \ie what skills for implementing the strategies inspired the participants, 
mentioned in their oral reports. 
\fan{In the specific task, a valid point had to describe at least one concrete implementation element of the strategy in the video, such as its timing, task design, or teacher behavior; merely restating the strategy definition was not counted.}
\fan{We treated sub-sentence phrasal segments as the basic coding unit.}
Furthermore, we invited a mathematics teacher (Male, 9 years experience) and a Chinese teacher (Female, 32 years experience) to evaluate the recorded videos of participants' teaching tests about their respective subjects. 
The evaluation process was fully blind, with raters unaware of the experimental condition for each video.
Based on the literature on microteaching~\cite{thangaraju2023microteaching, he2011exploring} and the scenario of VBL, we established the following four \textit{metrics of teaching performance} on 7-point Likert Scale (1/7 -- worst/best performance):
% [@Haoxiang, right?]}
% : 
a) instructional competence,
b) application of teaching strategies learned from videos (\eg choosing the right moment and appropriate way to introduce a strategy),
c) imitation of concrete implementations of the demonstrated strategies (\eg reproducing the teacher’s gestures or phrasing), and
d) the authenticity of the scenario.
}

% \subsubsection{Quantitative Assessment.}
% After using either the baseline system or \name{}, participants completed a questionnaire designed to quantitatively assess usability \& usefulness of the system (RQ1), and users' perception of effectiveness, inspired by prior work~\cite{lund2001measuring, pan2025tutorup,kirkpatrick2006evaluating}, as shown in~\autoref{tab:questionnaire-rq}.
% All questionnaire items were measured using a 7-point Likert scale (1 = strongly disagree, 7 = strongly agree). 
\zhh{
For \textbf{RQ2}, we adapted measures from~\cite{lund2001measuring, pan2025tutorup, kirkpatrick2006evaluating} (\autoref{sec:questionnaire}) and adopted 7-point Likert Scale (1/7 -- strongly disagree/agree) to measure the following items after each learning session: 
% \zhhred{[@Haoxiang, change the appendix and result tables accordingly]}
Q1) confidence in teaching,  
Q2) gained insights, 
Q3) benefits of practicing sessions, 
Q4) engagement in the learning process,
Q5) ease of use, and
Q6) intention to use.
}

\section{Analyses and Results}
\zhh{In this section, we present the quantitative and qualitative results for each research question.
\autoref{tab:all_results} summarizes all quantitative comparisons between \name{} and the baseline.}
\zhh{
For the numbers of valid inspiration points (RQ1) and items about usability and usefulness (RQ2), we employ the Wilcoxon signed-rank tests to compare the differences between the \name{} and baseline condition.  
% As for the \textbf{items measured on a 7-point Likert scale} about the usability and usefulness (RQ1), we employ the Wilcoxon signed-rank tests to compare the differences between the \name{} and baseline condition in the within-subjects study. The \textit{``valid inspiration point''} also uses the same method to compare.
As for the expert evaluation of each participant's performance in the test (RQ1) after using \name{} and the baseline, we first apply Z-score standardization to mitigate inconsistencies in scoring standards among different raters. To compare the conditions, we then assessed the normality of the score differences using the Shapiro-Wilk test. 
A paired t-test was used for normally distributed data; otherwise, the Wilcoxon signed-rank test was applied.
For the qualitative data, 
% consistent with the formative study, 
two authors applied Braun and Clarke's six-phase thematic analysis framework~\cite{guest2011applied} using a deductive approach,
with predefined themes of effectiveness (RQ1), and usability \& usefulness (RQ2). 
% We incorporate these qualitative findings in the following presentation of our results.
}

\begin{table}[t]
\centering
\small
\caption{Quantitative comparison between \name{} and Baseline. Q1--Q6 are 7-point Likert items; Valid Inspiration Points are raw counts. Expert-rated scores (Dim.\ 1--4) are Z-score standardized across raters. +: \(p < .1\), *: \(p < .05\), **: \(p < .01\).}
\label{tab:all_results}
\setlength{\tabcolsep}{2pt}%
\renewcommand{\arraystretch}{1.2}% approximates tabularray's rowsep = 1.2pt
\begin{tabularx}{\linewidth}{
  >{\hsize=0.40\hsize\centering\arraybackslash}X    % RQ   (must fit the \multirow "RQ1"/"RQ2" box)
  >{\hsize=2.71\hsize\raggedright\arraybackslash}X  % Item
  >{\hsize=1.10\hsize\centering\arraybackslash}X    % TeachUp Mean (SD)
  >{\hsize=1.10\hsize\centering\arraybackslash}X    % Baseline Mean (SD)
  >{\hsize=0.34\hsize\centering\arraybackslash}X    % p
  >{\hsize=0.35\hsize\centering\arraybackslash}X    % Sig.
}
\toprule
% ===== 表头 =====
\textbf{RQ} & \multicolumn{1}{c}{\textbf{Item}} & \textbf{\shortstack{TeachUp\\Mean (SD)}} & \textbf{\shortstack{Baseline\\Mean (SD)}} & \(p\) & \textbf{Sig.} \\
\midrule
% ===== RQ1: Reflection + Expert-rated Performance =====
\multirow{5}{*}{RQ1}
  & Valid Inspiration Points               & \fan{2.56 (1.09)} & \fan{1.81 (1.11)} & \fan{.049} & \fan{*} \\
\cmidrule(l){2-6}
  & Dim1: Instructional competence         & \(+\)0.27 (0.99) & \(-\)0.27 (1.00) & .065 & +  \\
  & Dim2: Strategy application             & \(+\)0.32 (0.95) & \(-\)0.32 (1.01) & .003 & ** \\
  & Dim3: Imitation of implementations     & \(+\)0.09 (1.22) & \(-\)0.09 (0.79) & .516 &    \\
  & Dim4: Authenticity of scenario         & \(+\)0.24 (1.13) & \(-\)0.24 (0.86) & .086 & +  \\
\midrule
% ===== RQ2: Questionnaire (Q1--Q6) =====
\multirow{6}{*}{RQ2}
  & Q1: Confidence in teaching             & 5.69 (0.79) & 5.00 (1.21) & .009 & ** \\
  & Q2: Gained insights                    & 5.88 (1.02) & 5.75 (0.86) & .527 &    \\
  & Q3: Benefits of practicing sessions    & 6.25 (0.58) & 5.63 (0.72) & .014 & *  \\
  & Q4: Engagement in the learning process & 6.13 (0.50) & 5.31 (1.20) & .026 & *  \\
  & Q5: Ease of use                        & 6.06 (0.77) & 5.63 (0.62) & .053 & +  \\
  & Q6: Intention to use                   & 6.13 (0.50) & 5.25 (1.18) & .017 & *  \\
\bottomrule
\end{tabularx}
\end{table}

\subsection{RQ1: Effectiveness of \name{}}
\zhh{
% As shown in \autoref{tab:all_results}, the number of valid inspiration points gained from the videos did not differ significantly ($p=.809$), suggesting both systems supported comparable observation of teaching techniques in the watching stage.
\fan{As shown in \autoref{tab:all_results}, participants identified significantly more valid inspiration points---that is, concrete implementation elements of the focal strategy---with \name{} ($M=2.56$, $SD=1.09$) than with the baseline ($M=1.81$, $SD=1.11$; $W=18.00$, $p=.049$).}
\textbf{Expert-rated performance} revealed that \name{} significantly improved participants' performance on applying the learned instructional strategies from the videos to a new teaching task ($p=.003$), with marginal improvements in instructional competence ($p=.065$) and scenario authenticity ($p=.086$). No significant difference was found in imitation of implementations of the strategies on the new teaching task ($p=.516$). %, consistent with the similar valid inspiration point counts. 
% Overall, \name{} is more effective at helping users translate observation into demonstrably improved teaching performance.
}

\zhh{Participants shared how \name{} supported the effectiveness of their learning. %, which is summarized below.
% \subsubsection{Effectiveness in Promoting Reflection in VBL}
In the video-watching stage, teachers regarded the AI-generated hints as a form of reflection for identifying and addressing gaps. Some participants (P3, P11, P5, P6, P8) preferred to watch the entire video before answering the questions, and they expressed that these hints helped them recall details they had missed, as P11 said, \textit{``There are indeed some teaching methods, or some teaching steps, that I overlooked. I think AI in this respect can really serve as a good reminder.''}
In the scenario-based practicing stage, almost all users agreed that the tasks were helpful and guided them to improve their performance (except P11). 
For example, participants shared comments such as, \textit{``It gave me a task to think about my own wording''} (P14); \textit{``The coherence phrases I learned in the first stage guided me to apply them here''} (P16); \textit{``It (the task) helped me grasp the pace of teaching''} (P15). 
Overall, participants expressed that the AI-generated tasks not only improved their performance from different aspects (\eg language, pacing) but also encouraged deeper reflection on how teaching methods could be approached and refined.
}

\subsection{RQ2: Usability and Usefulness of \name{}}
\zhh{As shown in \autoref{tab:all_results}, compared to the baseline system, participants with \name{} reported significantly higher confidence in classroom teaching after the learning session ($p=.009$). 
There is no significant difference between the two conditions regarding perceived gained insights from the video cases that participants can directly apply to improve their teaching practices ($p=.527$).
Nevertheless, participants with \name{} found that its practice sessions were significantly more beneficial for applying the teaching skills learned from the videos in classroom settings ($p=.014$). 
% participants perceived \name{} as having a significantly more positive impact on \textbf{self-reported outcomes}: teaching confidence was significantly higher ($p=.009$), and practice sessions were rated significantly more effective for applying teaching techniques ($p=.014$). Gained insights showed no significant difference ($p=.527$).
Overall, participants perceived that \name{} is generally easier to use ($p=.053$), reported significantly higher intention to use \name{} for video-based learning and teaching reflection ($p=.017$), and were more interested and engaged in the training process with \name{} ($p=.026$), compared to the baseline system. 
% We summarized the qualitative feedback on \name{} as below.
% Regarding \textbf{usability}, intention to use ($p=.017$) and engagement ($p=.026$) were both significantly higher, and ease of use showed a marginal advantage ($p=.053$) despite \name{} integrating more features.}
% \haoxiang{Participants shared their perceptions of \name{}'s usability and usefulness in the post-condition interviews. The qualitative findings are presented below.}
}

% \zhh{
% \subsubsection{Overall Usability Perception}
% % All participants appreciated the structured learning process in \name{}.% integration. 
% All participants in within-subjects study expressed their overall appreciation for the integration.
% Users were able to clearly understand the purpose behind combining these functions from their first use. For example, P1 described: \textit{``The first page (watching stage) provided us with a good sample, the second page (practicing stage) offered a valuable practice opportunity, ... and in the third part, I could roughly know whether what I said covered the points I needed to practice.''}
% \haoxiang{Regarding the efficiency of AI-generated content in our system, three users (P10, P13, P8) pointed out that excessively lengthy outputs can detract from the user experience. P13 noted: \textit{``The language could be more concise. Sometimes just providing a few keywords or key points is enough to understand, without displaying so much.''}}
% }

% \subsubsection{Usefulness of Specific Features}

% \textbf{Reflective hints in the video-watching stage.}
\zhh{
All participants appreciated \name{}'s reflective support in the structured three-stage learning process. 
In the watching stage, most participants found the reflective questions helpful for learning from the videos (except P2, P3, and P15). These hints were seen as a form of guidance, encouraging users to pay attention to the micro-level of teacher's actions in the videos, sometimes even just from a glimpse. 
As P9 stated: \textit{``The analysis is very clear, for example explaining step by step how things continue, which lets me understand how the teacher interacts with the students step by step.''} 
All participants except P11 and P15 also found that the tasks in the practicing stage were helpful. 
They emphasized the importance of maintaining their own subjectivity rather than strictly following the tasks, as P15 remarked, \textit{``When connecting to these tasks it gave, sometimes I would suddenly get stuck; in contrast, (without these tasks), I know what I need to do and which skills I can use to make the teaching smoother and more complete.''} 
We recognize this as a positive phenomenon, as it reflects the teachers' reflective engagement during the practice process.
% Lastly, all participants praised the precision of the system's evaluation on the practices, with P8 noting it accurately identified what they failed to accomplish and offered concrete positive examples to learn from. 
\fan{Lastly, participants praised the precision of the system's evaluation of their practices. P8 remarked, \textit{``Its analyses were very accurate; it truly pointed out what I failed to accomplish.''} Similarly, P14 appreciated that \textit{``It gave me a specific positive example to learn from.''} Together, these comments illustrate how the evaluation supported reflection by identifying unmet practice goals and providing a concrete reference for improvement.}
% In contrast, those who found the usefulness to be low indicated that having too many tasks limited the overall usefulness.
}

\zhh{
Apart from the benefits, participants also reported room for improving \name{}. 
For example, several participants mentioned that having too many tasks in the practicing stage limited \name{}'s usefulness. 
Three users (P10, P13, P8) pointed out that excessively lengthy generated outputs can detract from the user experience, as P13 noted, \textit{``The language could be more concise. Sometimes just providing a few keywords or key points is enough to understand, without displaying so much.''} 
P3 further noted that in the evaluative report, some spatial or gestural details about his practice video were missed.
}

% \textbf{Reflective tasks in the practice-evaluation stage.}
% Almost all participants found the tasks provided during practice to be useful, except for P11 and P15. They emphasized the importance of maintaining their own subjectivity rather than strictly following the tasks. As P15 remarked, \textit{``When connecting to these tasks he gave, sometimes I would suddenly get stuck; in contrast, (without these tasks), I know what I need to do and which skills I can use to make the teaching smoother and more complete.''} 
% Nevertheless, we recognize this as a positive phenomenon, as it reflects the teachers' reflective engagement during the practice process.

% \textbf{Evaluation in the practice stage.}
% \haoxiang{All participants praised the precision of the system's evaluation, with P8 noting it accurately identified what they failed to accomplish and offered concrete positive examples to learn from. However, some spatial or gestural details were occasionally missed, as noted by P3.}

\section{\zhh{Interviews with Teachers}}
\zhh{
Following the within-subjects study, we conducted an interview study with four in-service teachers who used \name{} on their  subjects of interest and video clips to enhance the generalizability of our findings.
% \subsection{Participants and Procedure}
Two of them are early-stage teachers with one year of teaching experience -- one teaches science in an elementary school (N1, male), and the other teaches English in a middle school (N2, female). 
% The participants included two novice teachers, each with one year of teaching experience: one teaching elementary science (N1, male) and the other teaching middle school English (N2, female).
The remaining two are experienced teachers, including one Chinese teacher with 32 years of experience in an elementary school (E1, female) and one mathematics teacher (E2, male) with 29 years of experience in a high school. 
% In addition, two experienced teachers participated: one teaching elementary-level Chinese (E1, female) with 32 years of experience, and the other teaching high school mathematics (E2, male) with 29 years of experience. 
}

% We then conducted a 60-minute interview session with each participant. 
% The session began with an introduction to the background of \name{}, along with the nine teaching strategies, to ensure participants fully understood our motivation and the system. This was followed by a 10-minute tutorial on how to use \name{}.
% Participants were then given 50 minutes to freely explore the system using a think-aloud protocol.
% We conducted brief, contextual interviews each time they transitioned between \name{}'s main stages (\ie after watching a video, after completing the practice, and after reviewing the evaluation) to capture their immediate feedback.
% They were first instructed to watch at least two instructional cases, select one they liked, read the hints in full, and attempt to answer the questions posed by the system. 
% Subsequently, they were required to complete the practice and evaluation sections, reviewing all hints and feedback provided. At each stage, participants were asked questions to assess the usefulness of individual components. At the end of the session, participants answered questions from the perspective of in-service teachers regarding the potential effectiveness of \name{}, and its contribution as a professional development method. Throughout the entire process, we recorded both the screen and audio for subsequent analysis.
\zhh{
% \textbf{Procedure}. 
The day before the interview, we asked each participant to provide a topic within their subject with which they were highly familiar.
We then created a video resource library for each topic by downloading five recorded classroom videos from Bilibili, segmenting them, detecting the instructional strategies using our pipeline, and storing them in the database.
Each interview lasted 60 minutes and began with a brief introduction to \name{} and the nine teaching strategies, followed by a 10-minute system tutorial. 
Participants then freely explored the system for 50 minutes using a think-aloud protocol, completing at least two instructional cases and the full watch--practice--evaluate cycle. Brief contextual interviews were conducted at each stage transition to capture immediate feedback. The session concluded with questions on the potential of \name{} to support teachers' professional development. 
We analyzed the recorded screens of system interactions and audio of interviews and reported the results below. 
% Screens and audio were recorded throughout for subsequent analysis.
}

\zhh{
% \subsubsection{Usefulness of the Detection Pipeline}
\textbf{The strategy detection pipeline is useful}. 
\haoxiang{Because in-service teachers could freely explore  tens of video clips segmented by the proposed pipeline, we mainly focus on the evaluation of the pipeline's usefulness in this interview study.}
The teachers agree that the video segmentation had high precision. E2 further praised the diversity across clips: \textit{``Although the clips covered the same teaching point, they unfolded from many different angles, so I could find the clips that best fit my own teaching situation.''}
}

\zhh{
\textbf{The reflective support is helpful in the whole learning session.}
Experts generally praised our ``watch-practice-evaluate'' framework, affirming that it could enhance teachers' general competencies within a single session. E1 emphasized that benefits extend beyond verbal expression to classroom organization, discipline, and the habit of self-reflection: \textit{``It can also cultivate the habit of reflecting on whether one's classroom has met its objectives.''} E2 similarly stressed that reflective thinking needs to be verified through practice to take effect, and valued the platform's ability to support such simulation at any time.
}

\zhh{\textbf{Experienced teachers hold different views on \name{}'s long-term effectiveness}}. 
Regarding the long-term effectiveness of early-stage teacher professional development, the two experienced teachers offered differing views. E2 highlighted that reflective hints could support teachers in developing a more comprehensive understanding of their classrooms: \textit{``In everyday teaching, there may be some fixed mindsets... By learning through the system, thinking can be expanded, which is very meaningful.''} In contrast, E1 questioned the system's impact on sustained competence development, especially concerning knowledge transfer: \textit{``There may be some shortcomings in transfer, because the whole evaluation is based on comparing with the original video. I think teachers may need to watch the corresponding videos for all their classes.''}

% \subsection{Results and Analysis}
\zhh{
% \textbf{Effectiveness and Professional Development}
\textbf{The practicing tasks are inflexible and could be risky.}
In-service teachers indicated that fixed tasks made the training less flexible and posed risks for applying the learned instructional strategies in the real classroom. First, they pointed out that overly detailed task descriptions might lead to the development of inappropriate classroom habits. E2 noted, \textit{``if each step is too detailed, it may affect the efficiency of classroom teaching.''} Second, they raised concerns about microteaching without students, relying solely on `imagined' students and imitation of the teachers in the videos carries certain risks. 
N2 worried: \textit{``If the students' foundation is weak, some of the steps I just practiced may not be feasible in reality.''} To address this issue, E1 emphasized the importance of simulating student feedback: \textit{``I think student feedback is very important. It should include not only these pre-generated things, but also the feedback from students on what you just taught. So I suggest adding simulated student feedback in the future.''} This indicates the need for additional features to provide teachers with a more authentic and customizable environment, enriched with synchronous feedback.
}

% \subsubsection{Usability of System Features}
% Old: three separate \textbf{} paragraphs on AI content quality, reflective hints, and evaluation
\zhh{
\textbf{The generated content can be further improved.}
% \haoxiang{Regarding the AI-generated content,} 
Similar to the findings in the user study, 
E1, N1, and N2 noted that excessively lengthy generated content can detract from the user experience. N2 suggested,  
% that optimizing the user interface could help mitigate the issue: 
\textit{``Maybe keywords could be bolded or specially highlighted, so users can quickly and clearly identify them at a glance.''}
% \haoxiang{For the reflective hints,} 
% E1 provided a more concrete explanation 
E1 commented on the reflective hints, based on her familiarity with the selected Chinese class: \textit{``Because the video is short, one task may already be enough... The questions are quite repetitive, meaning the same question keeps being asked again and again.''}
Besides, N1 noted that some of his actions might not have been fully analyzed in the evaluative report. 
% \haoxiang{For the evaluation, N1 noted that some actions might not have been fully captured by the multimodal evaluation, particularly due to spatial constraints.}
}
\section{Discussion}

\zhh{
\subsection{Empowering Scalable Video-Based Learning for Teachers' Professional Development}
% This paper introduces an effective LLM-powered pipeline for automatically segmenting classroom video clips and detecting the teachers' instructional strategies. % detecting  instructional strategies in recorded classroom videos.
\fan{As an enabling component of \name{}, we implemented an LLM-powered pipeline for automatically segmenting classroom video clips and detecting the teachers' instructional strategies.}
% \zhh{On our benchmark with 89 labeled video clips, our pipeline reaches an acceptable precision of 0.634 for detecting nine strategies, which is acknowledged by the in-service teachers in our interview study.}
\fan{On our benchmark with 89 labeled video clips, our pipeline reaches a precision of 0.634 for detecting nine strategies, and the in-service teachers in our interview study found the resulting clips usable for their own learning.}
% The pipeline's effectiveness is demonstrated not only by its high detection accuracy but also by positive user perception: the in-service teachers who participated in our interviews acknowledged the precision of the video segmentation. 
% This confirms the success of the analysis approach based on transcribed text.
% This technical contribution addresses the design challenges of video-based learning (VBL) platforms identified by~\citet{bates2016if},
\fan{This component addresses the design challenges of video-based learning (VBL) platforms identified by~\citet{bates2016if},}
who found that merely providing a large video library can overwhelm teachers; instead, teachers prefer short, practical segments that focus on specific instructional strategies and are ``ready-to-use''. 
Our pipeline directly meets this need by automatically extracting such content, thereby saving teachers' time in searching for valuable learning materials online~\cite{gaudin2015video}.
\fan{Given the pipeline's preliminary performance, teachers could first screen detected clips by ranking them according to confidence or adjust the confidence threshold to balance reliability and breadth.}
\fan{Concise clip summaries and evidence linked to precise video moments could then help them quickly inspect why a strategy was suggested and decide whether the clip meets their learning needs.}
Besides, as highlighted by an experienced teacher in the interview, the diversity of video segments extracted by our pipeline demonstrates its potential for wide-ranging applications as a standalone tool. \haoxiang{We designed this pipeline as a reusable, ``plug-and-play middleware'' that can be deployed between any video resource database and online learning platform. Its generic input--output interface---taking a classroom video and producing timestamped strategy clips with category metadata---enables integration into diverse educational systems without platform-specific modification.} This decoupled design means that our technical contribution holds independent value and can enhance numerous existing online teacher education platforms.
% To improve our strategy detection pipeline and extend our benchmark, future work could leverage 
}

\zhh{
\subsection{ ``Watch-Practice-Evaluate'': A Kolb-Inspired Framework for Reflective VBL} %Video-Based Learning
We developed \name{}, an interactive system that supports a three-stage \textit{``watching-practicing-evaluating''} cycle for VBL, based on our formative study with both early-stage and experienced teachers. %design requirements.
% Our research shows that \name{} is not only effective, but users also expressed a desire to use it long-term for continuous reflection.
% \name{}'s 
This cycle can be viewed as a concrete implementation of Kolb's Reflective Cycle~\cite{kolb2014experiential}, tailored to the VBL context.
% Kolb's cycle begins with \textit{``Concrete Experience''}, which in our framework, involves observing an expert teacher's instructional case. 
% For novice teachers, this acts as a substitute for direct experience, offering highly contextualized and detailed practical insights that effectively bridge the gap in their own experience.
% Subsequently, \textit{``Reflective Observation''} and \textit{``Abstract Conceptualization''} in Kolb's cycle are implemented through reflective hints embedded in the watching stage. 
% Specifically, the ``learner-led reflection questions'' encourage users to link their observations with personal experiences, prompting them to reinterpret instructional strategies and transform these reflections into new ideas.
% Finally, \textit{``Active Experimentation''} from Kolb's cycle is implemented in our ``microteaching'' practice session, where users apply newly acquired concepts and receive immediate, personalized feedback, thereby completing the learning cycle.
\zhh{
Specifically, observing a teacher's well-performed instructional case serves as \textit{``Concrete Experience''}, acting as a substitute for early-stage teachers' limited teaching experience. 
The reflective hints in the watching stage implement \textit{``Reflective Observation''} and \textit{``Abstract Conceptualization''}, encouraging users to link observations with personal experiences and transform them into new ideas. Finally, the ``microteaching'' practice session realizes \textit{``Active Experimentation''}, where users apply newly acquired concepts and receive immediate, personalized feedback, completing the cycle.}
% The results of our user study provide empirical support for the effectiveness of this  framework. 
% Expert evaluations show that through learning with \name{}, novice teachers were able to significantly improve their ability to envision more authentic teaching scenarios and performed better overall. 
% This indicates that our proposed three-stage cycle—starting with a video-based ``substitutive experience'' and embedded within a complete cycle of reflection and practice—enhances novice teachers' pedagogical knowledge. 
% In summary, we present both an effective implementation framework and empirical evidence to guide the design of reflective VBL systems.
\zhh{%The results of our user study provide empirical support for this framework. %Expert evaluations show that 
In our user study, \name{} improved participants' ability to envision authentic teaching scenarios and enhanced their overall performance, suggesting that this cycle effectively enhances early-stage teachers' pedagogical knowledge. 
We thus present both an effective implementation framework and empirical evidence to guide the design of reflective VBL systems.
}
}

\zhh{
\subsection{Implications for HCI Research}
Our findings lead to three implications for HCI research.
\fan{
\subsubsection{\fan{AI-powered learning tools should bridge observation and action.}}
\fan{The results show that \name{} led to significantly more implementation elements of the focal strategy than the baseline.
Also, the qualitative results indicate that the watching-stage interactions play a grounding role in the integrated framework, providing strategy anchors for later practice and comparison-based evaluation.}
\fan{While many existing tools focus on helping users better observe and annotate video content~\cite{ngoon2024classinsight, rich2009video}, our findings suggest that these observations should be connected to enactment through contextualized practice, structured rehearsal, and personalized feedback grounded in the watched content.
This perspective may generalize to other domains where people learn by watching experts' demonstrations, such as medical training and public speaking.}
}

% \textbf{Implication 1: AI-powered learning tools should serve as personalized reflection partners, not mere assessment providers.} Previous HCI research on teachers' PD often viewed tools primarily as interfaces for reading assessments or for supporting teachers in filling out evaluations~\cite{thompson2019teacher, wang2021practice}.
% Instead, we designed the interactive reflection tool as a ``personalized reflection partner'' for users, rather than merely a provider of assessments. This design addresses researchers' calls for tools that guide users through reflection step-by-step, instead of simply delivering conclusions~\cite{fan2025litlinker, zhang2025friction}.
% \haoxiang{Aligning with the design approach of \cite{fan2024lessonplanner}, our system ``experiences'' the entire VBL journey alongside the user by providing relevant context (\eg transcripts of expert teachers and reflective questions in the watching stage), thereby offering more personalized feedback.}

\subsubsection{Using LLMs and multimodal large models to transform passive video resources into active learning environments.} 
\name{} demonstrates that large models can be prompted to extract pedagogical structure from videos and generate contextualized practice and evaluation around them, turning static recordings into interactive learning cycles. 
This extends emerging work on in-context video assessments~\cite{monserrat2014ive} and interactive notes~\cite{zhao2025noteit} by enabling multimodal feedback grounded in video content.
However, our findings indicate that generated content can sometimes be verbose or insufficiently attuned to nonverbal and temporal nuances in teaching videos. To address this, future systems could employ video generation models to synthesize diverse, fine-grained video variants for richer long-term rehearsal~\cite{xu2025recorded}, and leverage more specialized video-oriented large models capable of capturing subtle nonverbal cues (\eg gestures, posture) and cross-time-span pedagogical features~\cite{kim2025speaking}.
}

\zhh{
\subsubsection{Connecting practicing tasks during VBL with real-world teaching scenarios}
In the practice stage, \name{} generates teaching scenarios and tasks based on the watched classroom video clip and user-input practice goal. 
While participants in the user study generally appreciated the practicing tasks and maintained their own subjectivity during practices, teachers in the interviews were concerned that the tasks were inflexible and risky. 
\fan{These concerns reveal a transfer-distance tension in connecting video-based practice with real-world teaching.}
\fan{If a practice task is too close to the original video context, users may reproduce the case rather than develop the ability to apply the underlying strategy flexibly; conversely, if the context is too different, early-stage teachers may struggle to transfer the newly learned strategy.}
% These concerns point out the need for future work to create a more realistic classroom teaching environment for practicing the instructional strategies learned from the videos. 
% \fan{Addressing this tension requires future work to create more realistic and adaptive classroom environments for practicing the instructional strategies learned from the videos.}
\fan{Future work could address this tension by creating more realistic and adaptive classroom environments for practicing strategies learned from videos.}
% For example, future work could code and model the students' behaviors in the classroom videos to simulate student agents~\cite{zhang2025simulating}. 
\fan{For example, future work could code and model the students' behaviors in classroom videos to simulate student agents with varied profiles and responses~\cite{zhang2025simulating,pan2025tutorup,chun2026argumath}.}
The VBL support tool can therefore offer a virtual classroom for teaching practices~\cite{king2022automated}, in which student agents can interact with teachers by raising hands, speaking, eye contact, and so on.
\fan{Future systems could further adapt how far each practice scenario departs from the watched video according to users' demonstrated understanding and performance.}
}

\subsection{Limitation and Future Work}
This study has several limitations. 
% First, the formative study interviewed teachers from only a limited range of subjects and grade levels and used just three low-fidelity prototypes to gather feedback, which may not be sufficient to derive generalizable design requirements. Conducting a brainstorming session or workshop could provide more valuable insights.
% Second, 
First, the participants in the within-subjects controlled lab study were not actual elementary Chinese or Math teachers but were prompted to act as them. This mismatch with specialized teaching subjects may impair their performance in the tests after the learning session. %Due to their limited content knowledge, they were rarely able to provide high-quality qualitative insights. % in the qualitative study. 
While our interview with in-service teachers helps generalize our findings, future research should involve more early-stage teachers using \name{} within their specialized subjects to evaluate its effectiveness and usability. 
% Although our interview study helped address this limitation, future research should involve more pre-service and in-service teachers using the system within their familiar subjects to obtain more perspectives on its effectiveness and usability. 
% Additionally, the small sample size (N=16) and the fact that only two teachers from were involved to evaluate participants' assessment videos make our quantitative findings less robust. 
% Furthermore, because the experiment was run consecutively, participants might have noticed differences between the baseline and our system, which could have affected their perceptions. Therefore, a between-subjects design with a larger sample size is required in the future.
% Third, if \name{} is regarded as a method for training teachers, 
Second, our study lacks an evaluation of user behavior (\eg whether their teaching behaviors changed after the training) and results (\eg whether their students demonstrated significant improvement) in their following teaching practices, %according to
which are important in Kirkpatrick's four levels of evaluation~\cite{kirkpatrick2006evaluating}. Future case studies could be conducted to comprehensively assess the impact of \name{} on teachers' professional development in the long term.
\fan{Third, the current baseline is not a component-level ablation, so our study evaluates the integrated reflective loop as a whole and does not isolate the effectiveness of individual interaction components.}
\zhhred{
% Third, our strategy detection pipeline is developed and validated upon a small size of benchmark with 89 video scripts. 
\fan{Fourth, our strategy detection pipeline was developed and validated on a small benchmark of 89 video scripts.}
% \fan{Moreover, the controlled study used manually annotated ground-truth clips, while the in-service teacher interviews did not identify any incorrectly tagged clips.}
\fan{The controlled study used manually annotated ground-truth clips, and the in-service teacher interviews did not identify any incorrectly tagged clips.}
\fan{Consequently, our current studies do not establish how errors in the pipeline would affect early-stage teachers' learning.
% Future work can use our benchmark and pipeline as starting points to iteratively improve them by using the pipeline to give initial labels to more video scripts, manually refining the labels, and updating the pipeline with more representative examples and multimodal video features.
Alongside examining the effects of such errors on early-stage teachers' learning, future work can use our benchmark and pipeline as starting points to iteratively improve them by using the pipeline to give initial labels to more video scripts and incorporating more multimodal video features.} 
%Fourth, the interactive design of our system was underdeveloped. Currently, we present AI-generated content, including hints and tasks, mainly as plain text with CSS styling. Moving forward, incorporating more interactive design elements (\eg highlighting key words to improve readability or using visual cues to guide users) has potential to enhance user satisfaction.
}

\section{Conclusion}
\zhh{
In this paper, we designed and developed an interactive system, \name{}, to support teachers to learn instructional strategies from videos of others' well-executed classroom lectures.
We implemented a computational pipeline powered by large language models to detect nine instructional strategies enacted by teachers in the short video clips segmented from the classroom videos. 
With these video clips, \name{} provides reflective hints and adaptive microteaching practices to engage teachers in a ``watch–practice–evaluate'' video-based learning cycle. 
Our within-subjects user study with 16 early-stage teachers and four interviews with in-service teachers demonstrated that \name{} can help teachers effectively learn instructional strategies and improve their instructional competencies. With these findings, our computational pipeline, and developed system, we offer practical implications to future video-based learning systems for teachers' professional development. 
}

\begin{acks}
\fan{
This work is supported by Guangdong S\&T Programme (2025B010 1120004) and General Projects Fund of the Natural Science Foundation of Guangdong Province in China grant (2024A1515012226).
}
\end{acks}

\bibliographystyle{ACM-Reference-Format}
\bibliography{references}


\clearpage
\onecolumn

\appendix
\label{sec:appendix}
\section{Marzano's Nine Events}
\autoref{tab:strategies} shows the definitions and the examples of the nine broad, high-probability categories of teaching strategies used in classroom.
\begin{table}[H]
\renewcommand{\arraystretch}{1.25}
\setlength{\extrarowheight}{2pt}
\caption{Instructional strategies and examples of implementation in the classroom}
\label{tab:strategies}
\begin{tabularx}{\linewidth}{@{} 
    >{\centering\arraybackslash\hsize=0.7\hsize}X 
    >{\raggedright\arraybackslash\hsize=1.1\hsize}X 
    >{\raggedright\arraybackslash\hsize=1.2\hsize}X 
@{}} 
\toprule
\multicolumn{1}{c}{\textbf{Strategy}} & \multicolumn{1}{c}{\textbf{Definition}} & \multicolumn{1}{c}{\textbf{Examples of Implementation}} \\ \midrule
\textbf{(1) Identifying Similarities and Differences} & Having students compare, contrast, and classify concepts. & - Use Venn diagrams to compare and classify items. \newline - Ask students to identify similarities and differences on their own, followed by group discussion. \\
\midrule
\textbf{(2) Summarizing and Note Taking} & Having students identify key information and restate it concisely in their own words. & - Summarize concepts that have already been taught. \newline - Encourage students to use a consistent format for note taking. \\
\midrule
\textbf{(3) Reinforcing Effort and Providing Recognition} & Connecting student effort to achievement and providing recognition for their work. & - Assign timed quizzes and have students report on their performance (\eg accuracy). \newline - Display exemplary student work in the classroom to recognize their effort. \\
\midrule
\textbf{(4) Homework and Practice} & Using assignments and practice to reinforce concepts and skills learned in class. & - Set aside class time for students to practice. \newline - Assign and review homework. \\
\midrule
\textbf{(5) Nonlinguistic Representations} & Representing and exploring concepts using visuals, physical models, and movement. & - Use drawings, videos, physical models, or role-playing to demonstrate concepts. \\
\midrule
\textbf{(6) Cooperative Learning} & Organizing students into groups to work collaboratively on academic tasks. & - Group students based on interests or common experiences and assign specific roles (\eg recorder, reporter). \newline - Implement activities such as group presentations. \\
\midrule
\textbf{(7) Setting Objectives and Providing Feedback} & Setting clear learning goals and providing students with timely, specific feedback. & - Continuously display unit goals on a whiteboard or in slides, and track progress to provide ongoing feedback. \\
\midrule
\textbf{(8) Generating and Testing Hypotheses} & Engaging students in proposing hypotheses and then designing tasks to test them. & - Ask students to predict what would happen if an aspect of a system were changed. \newline - Organize groups to complete a hands-on task and discuss the reasons for their success or failure. \\
\midrule
\textbf{(9) Cues, Questions, and Advance Organizers} & Using cues, questions, and organizers to activate prior knowledge before introducing new concepts. & - Before a new lesson, present an outline, graphic image, or mind map as an advance organizer. \newline - Pose key questions and pause after asking to encourage deeper thinking before students answer. \\ \bottomrule
\end{tabularx}
\end{table}

\section{The Slide Deck Used in the Formative Study}

\setcounter{figure}{4}
\begin{figure}[H]
\centering
\includegraphics[width=0.9\linewidth]{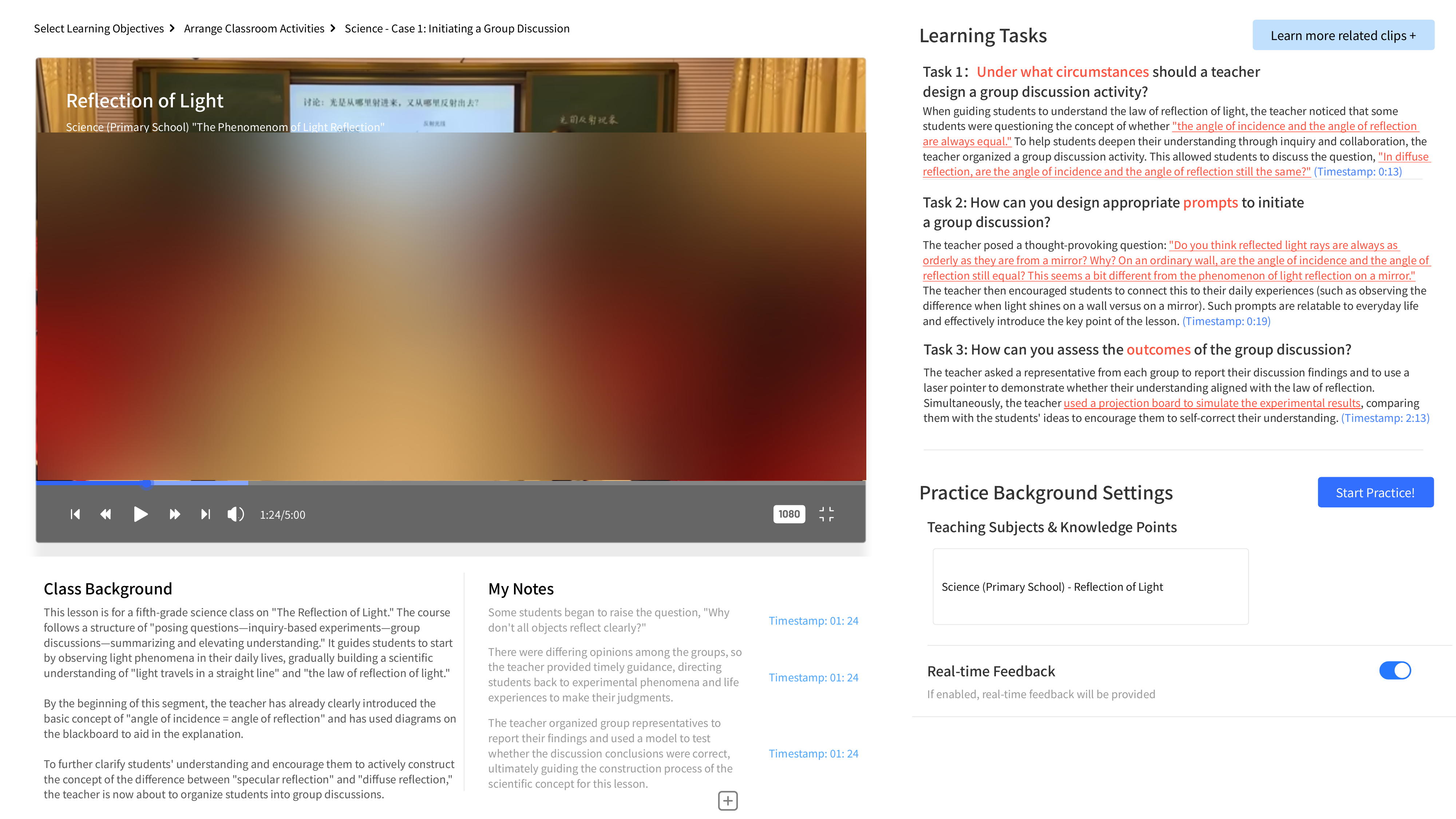} 
\caption{The video-watching page of the prototype. Translated from Chinese.
}
\Description{The video-watching page of the prototype. Translated from Chinese.}
\label{fig:proto1}
\end{figure}

\begin{figure}[H]
\centering
\includegraphics[width=0.9\linewidth]{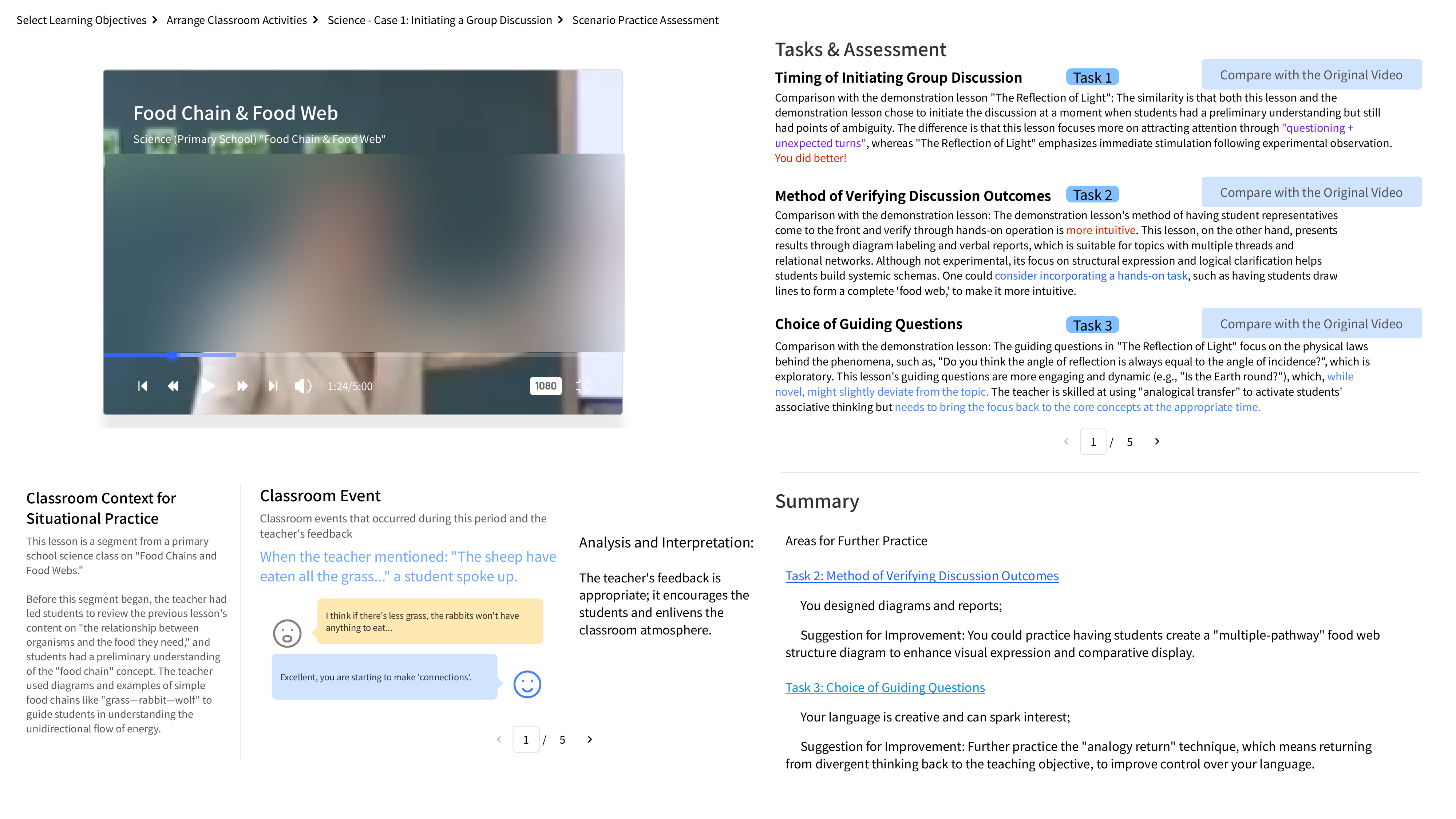} 
\caption{The evaluation page of the prototype. Translated from Chinese.
}
\Description{The evaluation page of the prototype. Translated from Chinese.}
\label{fig:proto2}
\end{figure}

\section{Questionnaire Items}
\label{sec:questionnaire}
The questionnaire items used to evaluate the system are shown below.
\begin{itemize}
\item \textbf{Q1}: I feel more confident about classroom teaching.
\item \textbf{Q2}: The video cases provided insights that I can directly apply to improve my teaching practice.
\item \textbf{Q3}: The practice sessions were beneficial for applying the teaching skills learned from the videos in classroom settings.
\item \textbf{Q4}: The training process kept me interested and engaged throughout.
\item \textbf{Q5}: I found the system easy to use.
\item \textbf{Q6}: I would like to use this system frequently for video-based learning and teaching reflection.
\end{itemize}

\section{Detailed Information of the Participants Involved}
In this section in appendix, the detailed information of the participants involved is listed below. Note: FE1 works at a Teacher Development Center, an institution under the local Education Bureau responsible for curriculum research, teacher training, and professional development.

% \begin{table}[H]
% \centering
% \begin{threeparttable}
% \caption{Experienced and early-stage teachers involved in the formative study}
% \label{tab:formative}
% \begin{tabular}{cccc}
% \toprule
% \textbf{ID} & \textbf{Gender} & \textbf{Background} & \textbf{Subjects} \\
% \midrule
% FE1 & Female & 38 years of teaching experience & --\tnote{1} \\
% FE2 & Female & 21 years of teaching experience & Chinese in elementary school \\
% FE3 & Female & 25 years of teaching experience & Mathematics in elementary school \\
% FE4 & Male   & 3 years of teaching experience  & Chinese in elementary school \\
% FE5 & Male   & 3 years of teaching experience  & Mathematics in elementary school \\
% \midrule
% FN1 & Male   & Normal university student & Chemistry in senior high school \\
% FN2 & Female & Normal university student & History in senior high school \\
% FN3 & Male   & Normal university student & Physics in junior high school \\
% FN4 & Male   & Normal university student & Music in senior high school \\
% \bottomrule
% \end{tabular}
% \begin{tablenotes}
%     \item[1] FE1 works at a Teacher Development Center, an institution under the local Education Bureau responsible for curriculum research, teacher training, and professional development.
% \end{tablenotes}
% \end{threeparttable}
% \end{table}
% threeparttable is not on ACM TAPS's accepted-package list; rewritten below
% without it (plain superscript + a \footnotesize note after the tabular).
\begin{table}[H]
\centering
% \begin{threeparttable}
\caption{Experienced and early-stage teachers involved in the formative study}
\label{tab:formative}
\begin{tabular}{cccc}
\toprule
\textbf{ID} & \textbf{Gender} & \textbf{Background} & \textbf{Subjects} \\
\midrule
% FE1 & Female & 38 years of teaching experience & --\tnote{1} \\
FE1 & Female & 38 years of teaching experience & -- \\
FE2 & Female & 21 years of teaching experience & Chinese in elementary school \\
FE3 & Female & 25 years of teaching experience & Mathematics in elementary school \\
FE4 & Male   & 3 years of teaching experience  & Chinese in elementary school \\
FE5 & Male   & 3 years of teaching experience  & Mathematics in elementary school \\
\midrule
FN1 & Male   & Normal university student & Chemistry in senior high school \\
FN2 & Female & Normal university student & History in senior high school \\
FN3 & Male   & Normal university student & Physics in junior high school \\
FN4 & Male   & Normal university student & Music in senior high school \\
\bottomrule
\end{tabular}
% \begin{tablenotes}
%     \item[1] FE1 works at a Teacher Development Center, an institution under the local Education Bureau responsible for curriculum research, teacher training, and professional development.
% \end{tablenotes}
% \end{threeparttable}

% {\footnotesize\raggedright\textsuperscript{1}FE1 works at a Teacher Development Center, an institution under the local Education Bureau responsible for curriculum research, teacher training, and professional development.\par}
\end{table}

% \begin{table*}[]
% \centering
% \caption{Participants involved in the within-subjects study. VBL stands for video-based learning, and PD stands for professional development.} 
% \label{tab:exp_design_detailed}
% \begin{tabular}{cccccc}
% \toprule
% ID & Gender & Age & Background & Freq. of VBL for PD & Freq. of LLMs Usage \\
% \midrule
% P1 & Male & 24 & Education major & Monthly & Daily \\
% P2 & Male & 24 &  4 months of teaching exp & Yearly & Daily \\
% P3 & Male & 20 & 1 year of teaching exp & Have used & Weekly \\
% P4 & Male & 23 & Education major & Monthly & Daily \\
% P5 & Male & 25 & Education major & Have used & Weekly \\
% P6 & Male & 25 & Certified teacher & Weekly & Weekly \\
% P7 & Female & 19 & Education major & Monthly & Weekly \\
% P8 & Male & 21 & - & Weekly & Daily \\
% P9 & Female & 20 & 1 year of teaching exp & Monthly & Daily \\
% P10 & Female & 21 & Education major & Never & Daily \\
% P11 & Male & 23 & 1 year of teaching exp & Never & Daily \\
% P12 & Female & 23 & 1 year of teaching exp & Yearly & Daily \\
% P13 & Female & 19 & nearly 1 year of teaching exp & Have used & Daily \\
% P14 & Female & 24 & 1 year of teaching exp & Weekly & Daily \\
% P15 & Female & 22 & - & Weekly & Daily \\
% P16 & Female & 22 & - & Weekly & Daily \\
% \bottomrule
% \end{tabular}
% \end{table*}

\begin{table}[h]
\centering
\caption{Participants involved in the within-subjects study. VBL stands for video-based learning, and PD stands for professional development. \fan{All participants were pre-service teachers from normal universities; P9--P16 additionally reported teaching experience, including internships.}}
\label{tab:exp_design_detailed}
\begin{tabular}{cccccc}
\toprule
ID & Gender & Age & Background & Freq.\ of VBL for PD & Freq.\ of LLM Usage \\
\midrule
P1  & Male   & 24 & Normal university student   & Monthly & Daily  \\
P2  & Male   & 23 & Normal university student   & Monthly & Daily  \\
P3  & Male   & 25 & Normal university student   & Rarely  & Weekly \\
P4  & Female & 19 & Normal university student   & Monthly & Weekly \\
P5  & Male   & 21 & Normal university student              & Weekly  & Daily  \\
P6 & Female & 21 & Normal university student   & Never   & Daily  \\
P7 & Female & 22 & Normal university student              & Weekly  & Daily  \\
P8 & Female & 22 & Normal university student              & Weekly  & Daily  \\
\midrule
P9  & Male   & 24 & 4 months of teaching exp.              & Yearly  & Daily  \\
P10  & Male   & 20 & 1 year of teaching exp.                & Rarely  & Weekly \\
P11 & Male   & 25 & 1 year of teaching exp.                & Weekly  & Weekly \\
P12 & Female & 20 & 1 year of teaching exp.                & Monthly & Daily  \\
P13 & Male   & 23 & 1 year of teaching exp.                & Never   & Daily  \\
P14 & Female & 23 & 1 year of teaching exp.                & Yearly  & Daily  \\
P15 & Female & 19 & 10 months of teaching exp.             & Rarely  & Daily  \\
P16 & Female & 24 & 1 year of teaching exp.                & Weekly  & Daily  \\
\bottomrule
\end{tabular}
\end{table}

\end{document}